\documentclass[10pt,a4paper]{scrartcl}

\usepackage{CLICdp}
\usepackage{CLICdp_definitions}
\usepackage{float}
\usepackage[disable]{todonotes} 
\usepackage{adjustbox}
\usepackage{sansmath}
\usepackage{multicol}
\usepackage{pgf-pie}
\usepackage{pgfplots, pgfplotstable}
\usetikzlibrary{positioning}
\usepackage{threeparttable}
\usepackage{multirow}
\usepackage{makecell}
\usepackage{etoolbox}
\usepackage[section]{placeins}
\usepackage{afterpage}
\robustify\bfseries
\usepackage{wrapfig}
\newcommand{\imgc}[1]{(image credit: #1)}
\newcommand{\imcl}{\imgc{CLIC}}
\newcommand{\imdp}{\imgc{CLICdp}}

\title{\hspace*{1cm} The Compact Linear e$^+$e$^-$ Collider (CLIC):\newline Accelerator and Detector}

\notitlestamp 

\date{\today}

\onbehalfof{ {\em Input to the European Particle Physics Strategy Update} \\ \vspace*{0cm} {\em on behalf of the CLIC and CLICdp Collaborations} \\ \vspace*{.4cm} {\large 18 December 2018} }  

\addauthor{Contact person: A.~Robson\thanks{aidan.robson@cern.ch}}{\institute{1}\institute{2}\\}
\addauthor{Editors: P.\,N.~Burrows}{\institute{1}\institute{3},}
\addauthor{N.~Catalan~Lasheras}{\institute{1},}
\addauthor{L.~Linssen}{\institute{1},}
\addauthor{M.~Petri\v{c}}{\institute{1},\\}
\addauthor{A.~Robson}{\institute{1}\institute{2},}
\addauthor{D.~Schulte}{\institute{1},}
\addauthor{E.~Sicking}{\institute{1},}
\addauthor{S.~Stapnes}{\institute{1},}
\addauthor{W.~Wuensch}{\institute{1}}

\addinstitute{1}{CERN, Switzerland}
\addinstitute{2}{University of Glasgow, United Kingdom}
\addinstitute{3}{University of Oxford, United Kingdom}

\DeclareSIUnit\years{years}
\DeclareSIUnit\days{days}
\abstract{The Compact Linear Collider (CLIC) is a \si{\TeV}-scale high-luminosity linear \epem collider under development by international collaborations hosted by CERN.
This document provides an overview of the design, technology, and implementation aspects of the CLIC accelerator and the detector.
For an optimal exploitation of its physics potential, CLIC is foreseen to be built and operated in stages, at centre-of-mass energies
of \SI{380}{\GeV}, \SI{1.5}{\TeV} and \SI{3}{\TeV}, for a site length ranging between \SI{11}{\km} and \SI{50}{\km}.
CLIC uses a two-beam acceleration scheme, in which normal-conducting high-gradient \SI{12}{\GHz} accelerating structures are powered
via a high-current drive beam. For the first stage, an alternative with X-band klystron powering is also considered.
CLIC accelerator optimisation, technical developments, and system tests have resulted in significant progress in recent years. 
Moreover, this has led to an increased energy efficiency and reduced power consumption of around \SI{170}{\mega\watt}
for the \SI{380}{\GeV} stage, together with a reduced cost estimate of approximately \mbox{\num{6} billion \si{CHF}}.
The detector concept, which matches the physics performance requirements and the CLIC experimental conditions, has been
refined using improved software tools for simulation and reconstruction. Significant progress has been made on detector
technology developments for the tracking and calorimetry systems.
The construction of the first CLIC energy stage could start as early as \num{2026} and first beams would be available by \num{2035},
marking the beginning of a physics programme spanning \SIrange{25}{30}{\years} and providing
excellent sensitivity to Beyond Standard Model physics, through direct searches and via a
broad set of precision measurements of Standard Model processes, particularly in the Higgs and top-quark sectors.}

\graphicspath{ {./logos/}{./figures/} }

\usepackage{cleveref}
\newlength{\abc}
\AtBeginDocument{%
\renewcommand{\ref}[1]{\mbox{\Cref{#1}}}

}

\begin{document}

\titlepage

\section{Introduction}

The Compact Linear Collider (CLIC) is a multi-\si{\TeV} high-luminosity linear \epem collider
under development by the CLIC accelerator collaboration~\cite{clic-study}.
It is the only mature multi-\si{\TeV} lepton collider proposal. 
CLIC uses a novel two-beam acceleration technique, with normal-conducting accelerating structures
operating in the range of \SIrange{70}{100}{\mega\volt/\meter}.
Detailed studies of the physics potential and detector for CLIC, and R\&D on detector technologies,
are carried out by the CLIC detector and physics (CLICdp) collaboration~\cite{clic-study}. 

The CLIC Conceptual Design Report (CDR) was published in \num{2012}~\cite{cdrvol1,cdrvol2,cdrvol3}. 
The main focus of the CDR was to demonstrate the feasibility of the CLIC accelerator at \SI{3}{\TeV}
and to confirm that high-precision physics measurements can be performed 
in the presence of particles from beam-induced background. 
Following the completion of the CDR, 
detailed studies on Higgs and top-quark physics, with particular focus on the first energy stage,
concluded that the optimal centre-of-mass energy for the CLIC first stage is $\roots=\SI{380}{\GeV}$.

\noindent \begin{minipage}{\linewidth}
  \begin{minipage}{0.46\textwidth}
\noindent As a result, a comprehensive optimisation study of the CLIC accelerator complex was performed, 
by scanning the full parameter space for the accelerating structures, and by using the luminosity, 
cost, and energy consumption as a gauge for operation at \SI{380}{\GeV} and \SI{3}{\TeV}. 
The results led to optimised accelerator design parameters for the proposed staging scenario, 
with operation at \SI{380}{\GeV}, \SI{1.5}{\TeV} and \SI{3}{\TeV}~\cite{StagingBaseline}.
The recently updated luminosities for each stage are given in~\ref{tab:clicstaging}. CLIC provides $\pm 80$\% longitudinal electron polarisation and proposes a sharing between the two polarisation states at each energy stage for optimal physics reach~\cite{Roloff:2645352}.
  \end{minipage}
  \hspace*{.75cm}
  \begin{minipage}{0.46\textwidth}
    \vspace*{-.75cm}
    \begin{table}[H] \centering
      \caption{Baseline CLIC energy stages and integrated luminosities, $\mathcal{L}_{\textrm{int}}$, for each stage in the updated scenario~\cite{Roloff:2645352}. \label{tab:clicstaging}}
      \begin{tabular}{SSS}\toprule
        {Stage} & {$\sqrt{s}$ [\si{\TeV}]} & {$\mathcal{L}_{\textrm{int}}$ [\si{\per\ab}]} \\
        \hline
        1 &  0.38{ (and 0.35)} &  1.0 \\
        2 &  1.5             &  2.5 \\
        3 &  3.0             &  5.0 \\
        \bottomrule
      \end{tabular}
    \end{table}
  \end{minipage}
\end{minipage}

This document summarises the current progress of the CLIC studies. 
There have been many recent achievements in accelerator design, technology development, and system tests,
described in detail in~\cite{ESU18PiP} and summarised in~\cite{ESU18Summary}.
Large-scale CLIC-specific beam tests have taken place, and 
crucial experience has been gained from the expanding field of Free Electron Laser (FEL) linacs and new-generation light sources. 
Together they provide the demonstration that all the CLIC design parameters are well understood and achieved in beam tests, confirming that the performance goals are realistic.
An alternative scenario for the first stage, where the accelerating structures are powered by X-band klystrons, has also been studied.
The implementation of CLIC at CERN has been investigated, 
including civil engineering, electrical networks, cooling and ventilation,
installation scheduling, transport, and safety aspects. 
Systematic studies have put emphasis on optimising cost and energy efficiency.

Since the completion of the CDR, the CLIC detector was further optimised
through a broad range of simulation studies,
resulting in the CLICdet detector design~\cite{CLICdet_note_2017, CLICdet_performance}.
In order to increase the angular acceptance of the detector,
the final focusing quadrupoles are now placed outside the detector in the accelerator tunnel. 
The software suite for simulation and event reconstruction was modernised and tuned for use with CLICdet.
Detector technology developments have focused on the most challenging aspects of the experiment, 
namely the light-weight silicon vertex and tracker system and the highly-granular calorimeters.
The detector R\&D activities have resulted in technology demonstrators,
showing that the required performance is already achievable or will be achieved in the next phase,
compatible with the CLIC timescale~\cite{ESU18RnD}.

\ref{sec:accelerator} provides an overview of the CLIC accelerator design and performance
at \SI{380}{\GeV} for both the two-beam baseline design and the klystron-based option.
It describes the path to the higher energies, \SI{1.5}{\TeV} and \SI{3}{\TeV},
and gives an overview of the key technology developments. 
\ref{sec:accelerator} also outlines key achievements from beam experiments and hardware tests,
providing evidence that the performance goals can be met. 
The present plans for the CLIC implementation, with emphasis on the \SI{380}{\GeV} stage, are given, 
as well as estimates of the energy consumption and of the cost for construction and operation. 
In~\ref{sec:detector} the CLIC detector and its performance results through simulation and event reconstruction are described.
Recent progress on detector technology developments is summarised. 

CLIC provides excellent sensitivity to Beyond Standard Model physics, through direct searches and via a
broad set of precision measurements of Standard Model processes, particularly in the Higgs and top-quark sectors. 
The physics potential at the three energy stages has been explored in detail~\cite{ClicHiggsPaper, ClicTopPaper, ESU18BSM} and
is presented in a separate submission to the European Strategy Update process,
`The Compact Linear \epem Collider (CLIC): Physics Potential'~\cite{ESU18physics}; this and 
supporting documents can be found at the following location: 

\begin{center}
\textbf{Supporting documents: \url{http://clic.cern/european-strategy}}
\end{center}

\section{CLIC Accelerator}
\label{sec:accelerator}
The current staged CLIC design is based on the concepts and R\&D results of the CDR~\cite{cdrvol1,cdrvol2,cdrvol3}
and on further studies that have since been carried out.
The design of the first energy stage, with a luminosity target of $\mathcal{L}=\SI{1.5e34}{\per\centi\meter\squared\per\second}$,
has been systematically optimised for cost~\cite{StagingBaseline}.
To limit the cost of the upgrades, the ability to reuse the components in higher-energy stages has been integrated into the design. 
The parameters for the different energy stages are given in ~\ref{t:scdup1}.

\begin{table}[b!]
\caption{Key parameters of the CLIC energy stages.}
\label{t:scdup1}
\centering
\begin{tabular}{l l l l l l}
\toprule
Parameter                  &   Symbol         &   Unit &    Stage 1 &   Stage 2 &   Stage 3 \\
\midrule
Centre-of-mass energy               & $\sqrt{s}$              &\si{\GeV}                                     & 380     & 1500          & 3000\\
Repetition frequency                & $f_{\text{rep}}$        &\si{\Hz}                                     & 50      & 50            & 50\\
Number of bunches per train         & $n_{\mathrm{b}}$                 &                                              & 352     & 312           & 312\\
Bunch separation                    & $\Delta\,t$             &\si{\ns}                                      & 0.5     & 0.5           & 0.5\\
Pulse length                        & $\tau_{\text{RF}}$      &\si{\ns}                                      & 244     &244            & 244\\
\midrule
Accelerating gradient               & $G$                     &\si{\mega\volt/\meter}                        & 72      & 72/100        & 72/100\\
\midrule
Total luminosity                    & $\mathcal{L}$           &\SI{e34}{\per\centi\meter\squared\per\second} & 1.5     & 3.7           & 5.9 \\
Luminosity above \SI{99}{\percent} of $\sqrt{s}$ & $\mathcal{L}_{0.01}$    &\SI{e34}{\per\centi\meter\squared\per\second} & 0.9     & 1.4           & 2\\
Total integrated luminosity per year& $\mathcal{L}_{\text{int}}$ &\si{\per\fb}                                  & 180     & 444           & 708 \\ 
\midrule
Main linac tunnel length                  &                         &\si{\km}                                      & 11.4    & 29.0          & 50.1\\
Number of particles per bunch       & $N$                     &\num{e9}                                      & 5.2     & 3.7           & 3.7\\
Bunch length                        & $\sigma_z$              &\si{\um}                                      & 70      & 44            & 44\\
IP beam size                        & $\sigma_x/\sigma_y$     &\si{\nm}                                      & 149/2.9 & $\sim$ 60/1.5 & $\sim$ 40/1\\
Normalised emittance (end of linac) & $\epsilon_x/\epsilon_y$ &\si{\nm}                                      & 900/20  & 660/20        & 660/20\\
Final RMS energy spread & & \si{\percent} & 0.35 & 0.35 & 0.35 \\
\midrule
Crossing angle (at IP)              &                         &\si{\mrad}                                    & 16.5    & 20            & 20 \\
\bottomrule
\end{tabular}
\end{table}

\begin{center}
\begin{figure}[t!]
\includegraphics[width=\textwidth]{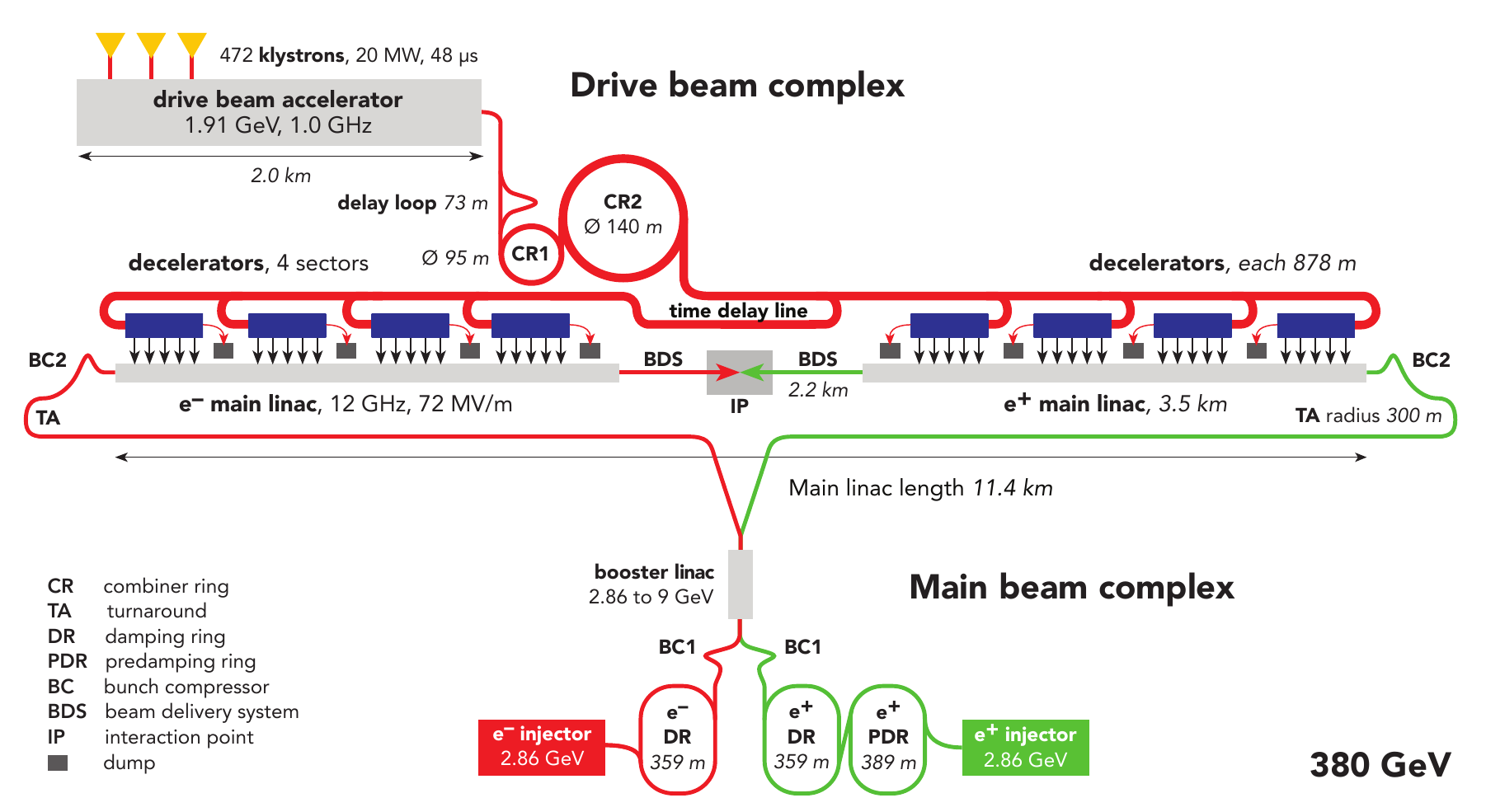}
\caption{Schematic layout of the CLIC complex at \SI{380}{\GeV}. \imcl}
\label{scd:clic_layout}
\end{figure}
\end{center}

\paragraph{Design and performance at 380\,GeV and extension to higher energies}
A schematic overview of the accelerator configuration for the first energy stage is shown in ~\ref{scd:clic_layout}.
The main electron beam is produced in a conventional radio frequency (RF) injector, which allows polarisation. The beam emittance is then
reduced in a damping ring.
To produce the positron beam, an electron beam is accelerated to \SI{5}{\GeV} and sent into a crystal to produce energetic photons,
which hit a second target and produce electron--positron pairs.
The positrons are captured and accelerated to \SI{2.86}{\GeV}.
Their beam emittance is reduced, first in a pre-damping ring and then in a damping ring.
The ring to main linac system accelerates both beams to \SI{9}{\GeV}, compresses their bunch length, and delivers the beams to the main linacs. 
The main linacs accelerate the beams to the collision energy of \SI{190}{GeV}.
The beam delivery system removes transverse tails and off-energy particles with collimators and compresses the beam to
the small size required at the interaction point (IP).
After collision, the beams are transported by the post collision lines to their respective beam dumps.

To reach multi-TeV collision energies in an acceptable site length and at affordable cost, 
the main linacs use normal conducting X-band accelerating structures;
these achieve a high accelerating gradient of \SI{100}{\mega\volt/\meter}.
For the first energy stage, a lower gradient of \SI{72}{\mega\volt/\meter} is the optimum to achieve the luminosity goal, which
requires a larger beam current than at higher energies.

In order to produce and support high gradients, the accelerating structures are required to be fed by short,
very high power RF pulses, which are difficult to generate at 
acceptable cost and efficiency using conventional klystrons.
In order to provide the necessary high peak power, the novel drive-beam scheme uses low-frequency klystrons
to efficiently generate long RF pulses and to store their energy
in a long, high-current drive-beam pulse.
This beam pulse is used to generate many short, even higher intensity pulses that are distributed alongside the main linac,
where they release the stored energy in power extraction and transfer structures (PETS) 
in the form of short RF power pulses, transferred via waveguides into the accelerating structures. 
This concept strongly reduces the cost and power consumption compared with powering the structures directly by klystrons.

\begin{figure}[b!]
\centering
\begin{subfigure}{.49\textwidth}
\subcaptionbox{\label{fig:IMP_1}}{\includegraphics[width=\textwidth]{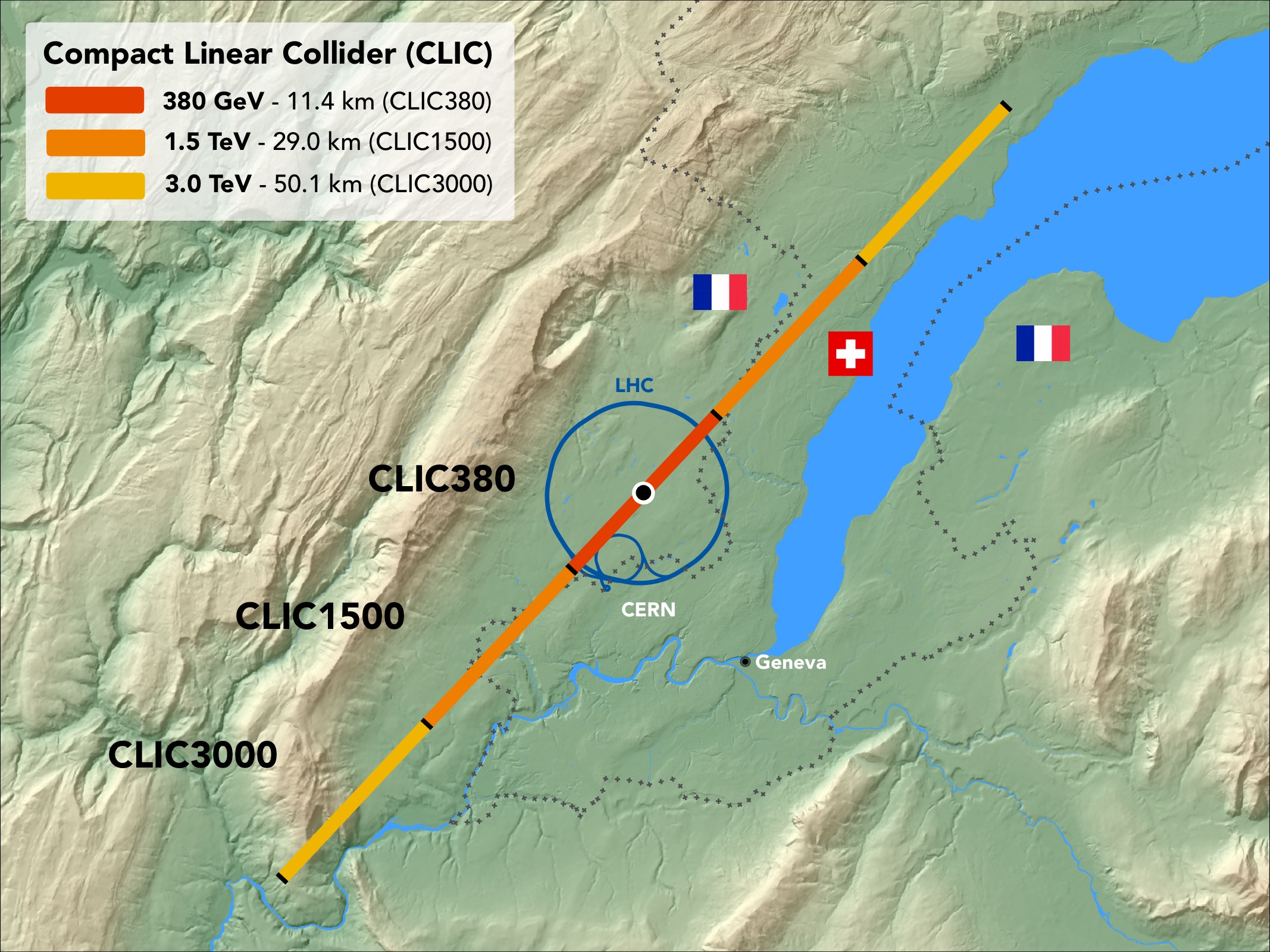}} 
\end{subfigure}
\hspace*{1cm}
\begin{subfigure}{.3\textwidth}
\subcaptionbox{\label{fig:CEIS_7a}}{\includegraphics[height=1.25\textwidth]{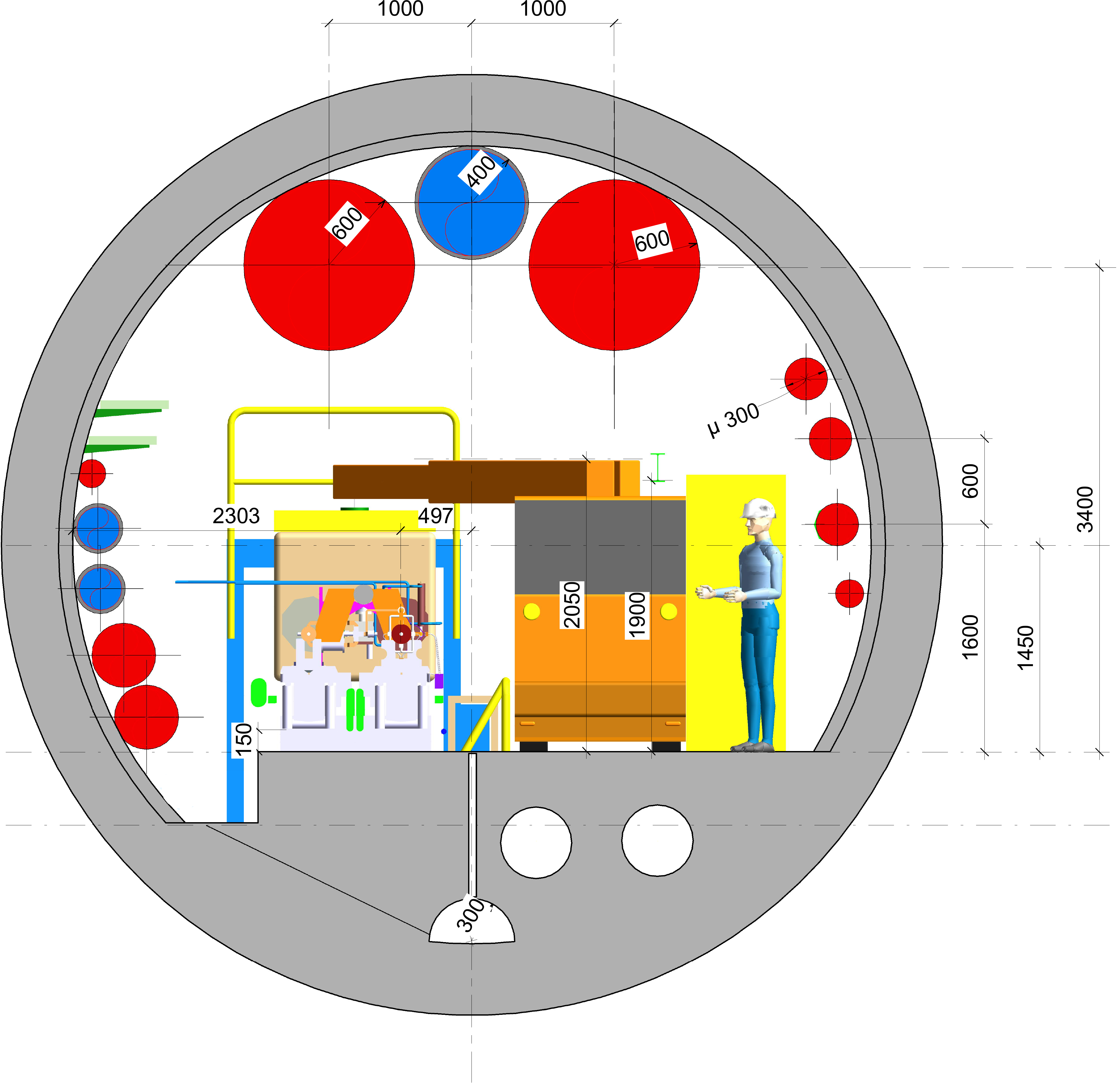}}
\end{subfigure}
\hspace*{1.5cm}
\caption{(a) The CLIC main linac footprint near CERN, showing the three implementation stages. (b) The CLIC main linac tunnel cross section for the drive-beam design. \imcl}
\end{figure}

The upgrade to higher energies involves lengthening the main linacs:  connecting new tunnels to the existing tunnels, 
moving the existing modules to the beginning of the new tunnels, and adding new, higher-gradient modules. 
When upgrading to \SI{1.5}{\TeV}, the length of the beam delivery system (BDS) needs to be increased and new magnets installed.
For the BDS upgrade to \SI{3}{\TeV}, only new magnets are required.
The central main-beam production complex needs only minor modifications for the first upgrade, to adjust to the smaller
number and charge of the bunches, and needs no further modifications for the second upgrade. 
For the upgrade to \SI{1.5}{\TeV}, the central drive-beam complex needs to be slightly extended to increase the drive-beam energy.
For the upgrade to \SI{3}{\TeV}, a second drive-beam complex must be added. 
This staged collider can be implemented at CERN, as shown in~\ref{fig:IMP_1}.
The main-beam and drive-beam production facilities are located at the CERN Pr\'evessin site and the
tunnel of the first two energy stages would be fully in molasse, which is ideal for tunnelling.
The main linac tunnel cross section is shown in~\ref{fig:CEIS_7a}.

The CLIC beam energy can be adjusted to meet different physics requirements. In particular, a period of operation around \SI{350}{GeV} is foreseen to scan the top-quark pair-production threshold.
Operation at much lower energies can also be considered. At the Z-pole, between \SI{2.5}{\per\fb} and \SI{45}{\per\fb} can be achieved per
year for an unmodified and a modified collider, respectively. 

\vspace*{-1cm}

\paragraph{Klystron-based alternative}
While the drive-beam scheme is optimal for the high-energy stages,
an alternative design for the \SI{380}{\GeV} stage has been considered, in which the main linac
accelerating structures are directly powered by klystrons.
This increases the cost per GeV in the main linac but it removes the cost of the drive-beam generation complex.
This means that it could be
a competitive alternative at lower energies provided that the cost of X-band klystrons and their modulators were sufficiently reduced.
The klystron-based design is very similar to the drive-beam baseline.
It requires a larger main linac tunnel that is separated into two parts by a shielding wall.
One part contains the accelerator; the other, the klystrons and related equipment. 
The collider can be upgraded by extending the main linacs with new sections that are powered by a drive beam.

\begin{figure}[htb]
\begin{center}
  \centering
  \includegraphics[width=0.75\textwidth]{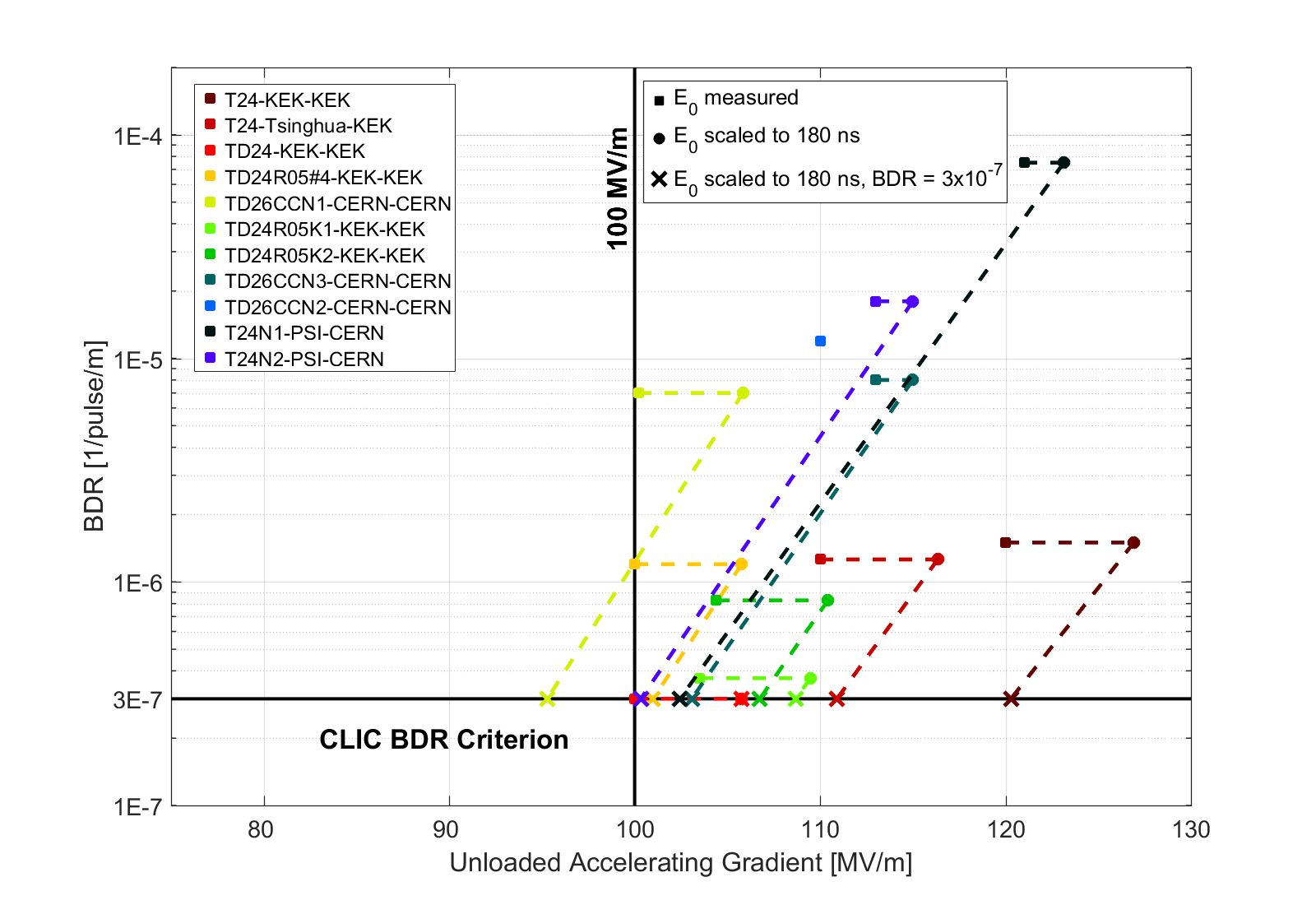}
 \caption{A summary of achieved performances of \SI{3}{\TeV} acceleration structures in tests.
 The vertical axis represents the breakdown rate (BDR) per metre.
 The final operating conditions of the tests are indicated by squares.
 Known scaling is used to determine the performance for the nominal CLIC pulse duration
 (dashed lines connecting squares to circles) and subsequently for the CLIC-specified breakdown rate of
 \SI{3e-7}{\per\meter} per pulse (dashed lines connecting circles to crosses).  Ten prototype
 structures can be seen operating above the target gradient of \SI{100}{\mega\volt/\meter}. \imcl
 }
\label{fig:RFstructure-test-results}
\end{center}
\end{figure}

\paragraph{Performance of the drive-beam concept}
The focus of accelerating structure development has been on achieving a gradient of \SI{100}{\mega\volt/\meter}
as required for the second and third energy stages.
The complex phenomena that occur at such high field levels, especially vacuum breakdown, have been understood,  
paving the way for the development of an RF design methodology,
fabrication techniques, and operational procedures that reduce the breakdown likelihood; 
the target is to have a rate of less than \SI{3e-7}{\per\meter} per beam pulse.
This ensures that only 1\% of the beam pulses experience a breakdown anywhere along the linacs, even at the \SI{3}{\TeV} stage.
Over a dozen prototype accelerating structures, with parameters consistent with beam dynamics
requirements, have now been operated at gradients above \SI{100}{\mega\volt/\meter} for extended periods
in dedicated test stands~\cite{ESU18PiP}.
A summary of the performances achieved in recent tests is shown in~\ref{fig:RFstructure-test-results}.

To test the drive-beam concept, the CLIC Test Facility (CTF3)~\cite{Geschonke2002} was constructed and operated by an international collaboration.
The hardware and experiments addressed the key points of the two-beam concept.
This included the demonstration of the high transfer efficiency from the RF to the drive beam and the stable drive-beam acceleration, 
as well as the generation of the final short, high-current pulse using a delay loop and combiner ring.
It also	assessed the final beam quality.
In particular, feedback was used to stabilise the drive-beam current and phase to ensure correct main-beam acceleration.
CTF3 achieved the drive-beam phase stability that is required for CLIC~\cite{c:phasetol,c:chetan_gm,c:phasetest}.
Finally, the drive beam was used to accelerate the main beam with a maximum gradient of \SI{145}{\mega\volt/\meter}.
This established the feasibility of the drive-beam concept and its use to accelerate a main beam.

\begin{wrapfigure}{r}{0.5\textwidth}
\begin{center}
\vspace{-0pt} \includegraphics[width=\linewidth]{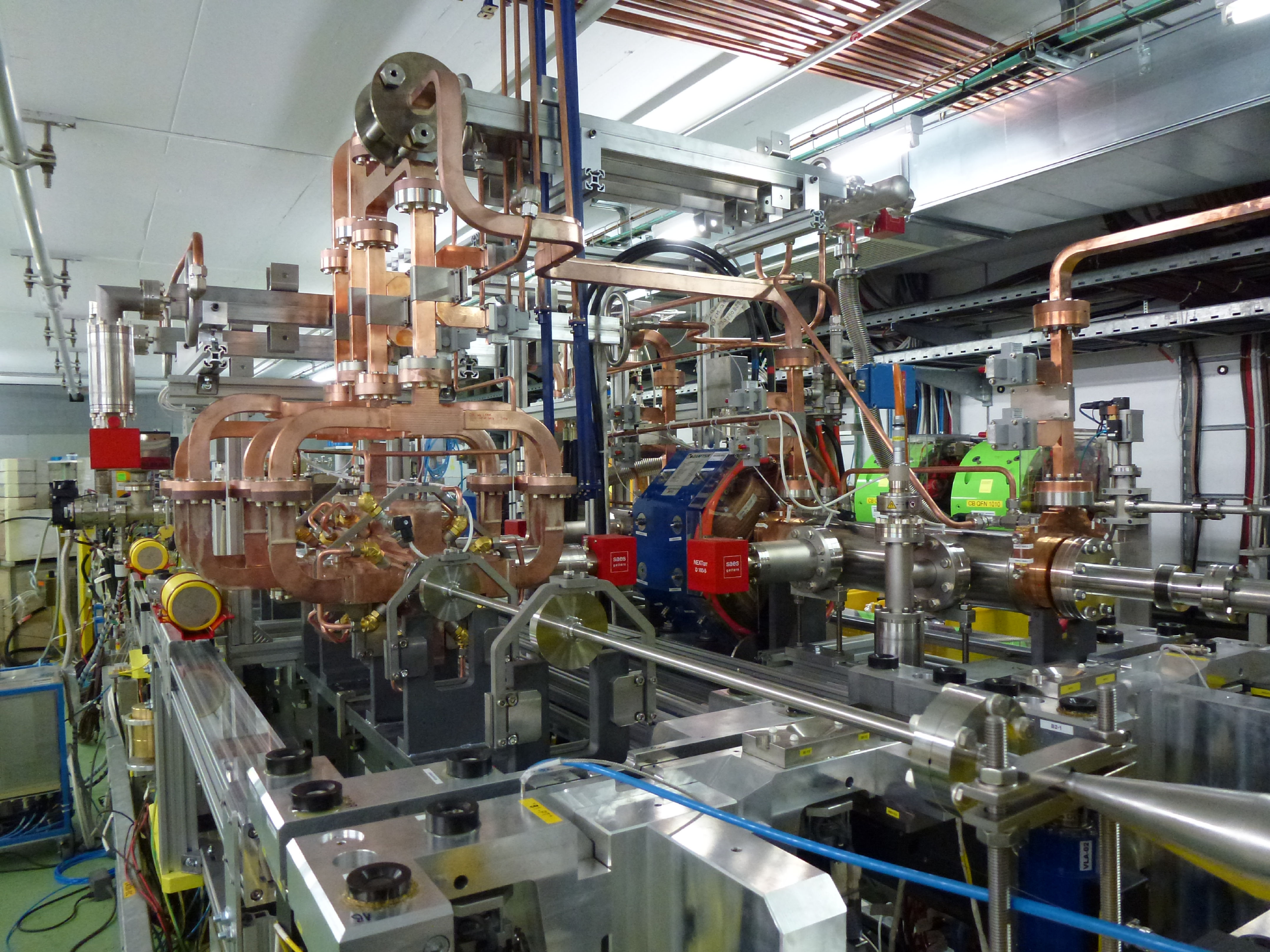}
\caption{The two-beam acceleration test stand in the CTF3 facility. The drive beam enters from the middle-right, while the probe (main) beam enters from the bottom-right. \imcl}
\label{fig:CTF3photo}
\end{center}
\vspace{-25pt}
\end{wrapfigure}

CTF3 has also been instrumental in developing and testing many different hardware components that are essential for the two-beam acceleration
scheme and
in validating their performance. Among them are the drive-beam gun, the bunch compressor, drive-beam accelerating structures,
RF deflectors, PETS including a power-off mechanism, the power distribution waveguide system, fast-feedback systems, and drive-beam current
and phase monitors.~\ref{fig:CTF3photo} shows the two-beam test stand that integrates all the critical components for the
drive-beam deceleration, main-beam acceleration process.
CTF3 stopped operation after successfully completing its experimental programme in December 2016 and a new facility,
CLEAR, has started to operate.  CLEAR re-uses the
CTF3 main-beam installations and additional hardware for further beam dynamics studies, with a focus on the main beam.

\paragraph{Luminosity performance}
The baseline plan for operating CLIC includes a yearly shutdown of 120 days, 30 days of commissioning,
20 days for machine development, and 10 days for planned technical stops.
This leaves 185 days of operation for collider physics data-taking. Assuming an availability during normal running of 75\%, this results in an integrated
luminosity per year equivalent to operating at full luminosity for \SI{1.2e7}{\second}~\cite{Bordry:2018gri}. 

In order to achieve high luminosity, CLIC requires very small beam sizes at the collision point, as listed in~\ref{t:scdup1}.
The resulting high charge density leads to
strong beam--beam effects, which result in the emission of beamstrahlung and production of background particles. This is limited to an acceptable level by
using flat beams, which are much larger in the horizontal than in the vertical plane. The key to high luminosity lies in the
small vertical beam size, and therefore a small vertical emittance and strong vertical focusing are essential.

The vertical emittance and consequently the luminosity are to a large extent determined by
imperfections in the accelerator complex; without them
a luminosity of $\mathcal{L}=\SI{4.3e34}{\per\centi\meter\squared\per\second}$ would be reached at $\roots=\SI{380}{\GeV}$.
The nominal luminosity of $\mathcal{L}=\SI{1.5e34}{\per\centi\meter\squared\per\second}$
takes into account the impact of the different imperfections; significant margin has been added to the known effects to enhance the robustness of the design.
Important examples of imperfections are the misalignment of components after the installation and the vibrations of the quadrupoles due to ground motion. 

Novel system designs and technologies in combination with beam-based tuning
minimise the effect of imperfections, and simulated performances provide full luminosity at the \SI{3}{TeV} stage.
While this is not strictly necessary for the initial energy stage, where the performance specifications could be relaxed,
this choice avoids the need for these systems to be upgraded for the higher-energy stages.
Key technologies that have been developed include the pre-alignment and quadrupole stabilisation systems
as well as high-precision beam instrumentation.
Pre-alignment at the
\SI{10}{\micro\meter}-level is required for the main linac and BDS components.
An active alignment system achieves this, using actuators and sensors to remotely align the
\SI{2}{\meter}-long modules that support the main linac components 
with respect to a stretched-wire reference network. The alignment system has been successfully tested.
The disks that make up the accelerating structures are fabricated in
industry to micron tolerance and are bonded together to a tolerance of ten microns.
To mitigate the effect of wakefields caused by misalignments, the accelerating structures are equipped
with monitors that measure this wakefield to
determine the offset of the structure from the beam. The remote alignment system uses this information to minimise the
wakefield effects on the beam.
A nm-level vibration stabilisation system for both main linac and final focus quadrupoles,
which is based on actuators and inertial sensors, allows decoupling of the focusing magnets from the ground.
This is important to avoid the natural ground motion as well as vibrations induced by
technical equipment that could lead to beam jitter and reduced luminosity.
Ground motion has been measured at CERN, and CLIC is designed to withstand the 
noise that has been measured in the CMS detector cavern.
Other key technologies that have been developed include hybrid, high-gradient final quadrupoles,
and advanced wigglers for the damping rings. 

In addition to the hardware tests, beam experiments provide the evidence that CLIC can reach its performance goals.
The Swiss Light Source and the Australian Light Source have reached CLIC-level vertical emittances~\cite{c:ls1, c:ls3}.
The advanced beam-based alignment of the CLIC main linac has successfully been tested in FACET and FERMI~\cite{FACET, FERMI}.

\paragraph{Other technology developments}
Further technology developments include the main linac modules and their auxiliary sub-systems
such as vacuum, stable supports, and instrumentation. 
Many developments have focused primarily on cost and power reduction~\cite{ESU18PiP}.
In both the drive-beam and klystron-based versions of CLIC, all of the power going to the beam
flows through klystrons.  Their efficiency is thus crucial to the overall efficiency of the facility.
Recently, new klystron beam dynamics concepts have been established, 
in studies driven by the CLIC collaboration, which result in significant efficiency increases
over a broad range of frequencies and power levels.  The concepts have been successfully demonstrated in a test of a retrofitted
S-band klystron.  The concepts are now being implemented in both the L-band drive-beam klystrons and in the X-band main linac klystron designs for CLIC, 
as well as klystrons for other projects that require high RF production efficiency.
CLIC requires several thousand focusing quadrupoles in the drive beam.
Adjustable permanent magnet prototypes have been developed
that reduce the power needs as well as the power supply and cabling costs.

Beam instrumentation, including sub-micron level resolution beam-position monitors
with time accuracy better than \SI{20}{ns} and bunch-length monitors with resolution better than \SI{20}{fs},
have been developed and tested with beam in CTF3. 
CLIC technology developments have also had a strong influence on small-scale,
compact accelerator applications from many fields.
The modularity of CLIC has the consequence that much of the linac technology,
especially that relating to high-gradient acceleration, is directly relevant for photon sources,
medical linacs and beam manipulation devices.  The adoption of CLIC technology for these
applications is now providing a significant boost to CLIC,
especially through an enlarging commercial supplier base~\cite{DAuria:2062591}. 

\begin{figure}[h!]
\centering
\includegraphics[width=\textwidth]{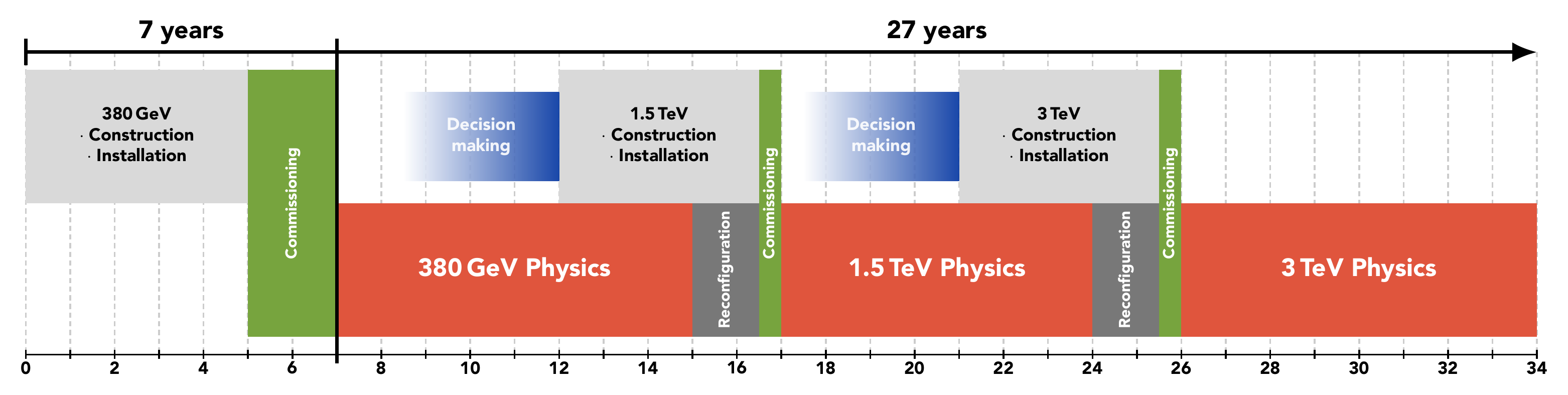}
\caption{\label{fig_IMP_9} Technology and construction-driven CLIC schedule, showing the construction and commissioning period and the three stages for data taking.
The time needed for reconfiguration (connection, hardware commissioning) between the stages is also indicated. \imcl}
\end{figure}

\paragraph{Construction, cost estimate, and power consumption}
The technology and construction-driven timeline for the CLIC programme is given in~\ref{fig_IMP_9}~\cite{ESU18PiP}.
This schedule has seven years of initial construction and commissioning,
potentially starting in 2026.  Including a two-year margin this leads to first collisions in 2035.
The 27 years of CLIC data-taking include two intervals of two years between the stages.

\vspace*{.7cm}
\vspace*{.7cm}

The cost estimate of the initial stage is approximately \num{5.9}~billion~\si{CHF}.  The energy upgrade
to \SI{1.5}{\TeV} has an estimated cost of approximately \num{5.1}~billion~\si{CHF}, including the upgrade of the
drive-beam RF power.  The cost of the further energy upgrade to \SI{3}{\TeV}
has been estimated at approximately \num{7.3}~billion~\si{CHF}, including the construction of a second
drive-beam complex.

The nominal power consumption at the \SI{380}{\GeV} stage is approximately \SI{170}{MW}.
Recent power savings at the \SI{380}{\GeV} stage have not yet been scrutinised for the
\SI{1.5}{\TeV} and \SI{3}{\TeV} stages, but earlier estimates give
approximately \SI{370}{MW} and \SI{590}{MW}, respectively~\cite{cdrvol1}.
The annual energy consumption for nominal running at the initial energy stage
is estimated to be \SI{0.8}{TWh}.  For comparison, CERN's current energy consumption is
approximately \SI{1.2}{TWh} per year, of which the accelerator complex uses approximately 90\%.

Further details on the schedule, cost estimate, and power consumption, including for the klystron-based option, 
are given in the Addendum and in~\cite{ESU18PiP}.

\clearpage
\section{CLIC detector design, technologies and performance}
\label{sec:detector}
The CLIC detector is a general-purpose detector designed to provide excellent performance for known
\epem physics processes as well as being prepared for a broad range of potential
Beyond Standard Model (BSM) signatures.
Its design and technology choices are driven by the CLIC physics
objectives and the experimental conditions. In this section, the requirements and design of
the CLIC detector are summarised, followed by a description of the detector technology choices and
the status of the detector R\&D. Examples of the detector performance are also given. 

In view of the physics objectives, the required track momentum resolution is 
$\sigma_{\pT}/\pT^2 \leq \SI{2e-5}{\per\GeV}$ for high-energy charged particles in the central detector region,
needed \eg for the precision measurement of heavy states decaying into leptons.
For optimal flavour tagging a resolution of \SI{5}{\um} in the transverse impact parameter, $d_0$, 
is required for single tracks in the central detector that have \pT above a few GeV. 
The corresponding limit on material thickness is \SI{0.2}{\percent}\,$\xo$ per layer in the vertex
detector and \SI{2}{\percent}\,$\xo$ per layer in the tracker. A single hit position resolution of $3\,\upmu\text{m}$
is required in the vertex detector and \SI{7}{\micro\meter} in the tracker.
The jet energy resolution, $\sigma_{E}/E$, is required to be better than \SI{5}{\percent} for light-quark jets
of  \SI{50}{\GeV}, and better than \SI{3.5}{\percent} for jet energies above \SI{100}{\GeV}.
This is needed, for example, to separate \PW, \PZ and \PH hadronic decays. Accurate jet reconstruction through
particle flow analysis (PFA) is at the origin of the choice of highly-granular sampling calorimetry.
Detector coverage for electrons and photons is needed down to small polar angles and \SI{10}{\mrad} is targeted.

As shown in~\ref{t:scdup1}, beams at CLIC arrive in bunch trains of \SIrange{156}{176}{\ns} duration.
During each bunch train at most one hard \epem collision is expected, while background particles
from beamstrahlung will be spread in time along the bunch train. The low event rates and the CLIC time structure allow for
triggerless readout of full bunch trains (at 50\,Hz). It also allows the on-detector power to be turned off during the
inter-train periods (power pulsing), thereby reducing the cooling needs. Beam-induced
backgrounds increase strongly with centre-of-mass energy, and so operation at \SI{3}{\TeV} can be considered 
the most challenging case for CLIC. The presence of beam-induced backgrounds imposes a hit time
resolution of \SI{1}{\ns} in the calorimeters and \SI{5}{\ns} in the tracking system. The design of
the CLIC detector foresees small readout cells in order to limit occupancies to at most 3\% integrated
over the bunch train. This leads to a maximum cell size of \SI{25x25}{\um} in the vertex detector and
maximum cell sizes of \SIrange{0.05}{0.5}{\mm^2} in the silicon tracker.
The small readout cells in the
calorimeters will allow not only for accurate jet reconstruction through PFA, but
will also provide an indispensable means of recognising particles from beam-induced background through
PFA reconstruction, and rejecting those particles by applying cuts on their \pT\ and arrival time. Except
for the very forward regions, radiation levels at CLIC are a factor of about~$\sim10^{-4}$ lower
than at LHC.

\begin{figure}[htb]
\centering
\begin{subfigure}{.5225\textwidth}
  \centering
   \subcaptionbox{\label{fig:clicdet}}{\includegraphics[width=\linewidth,trim={11cm 4.5cm 10cm 7cm},clip]{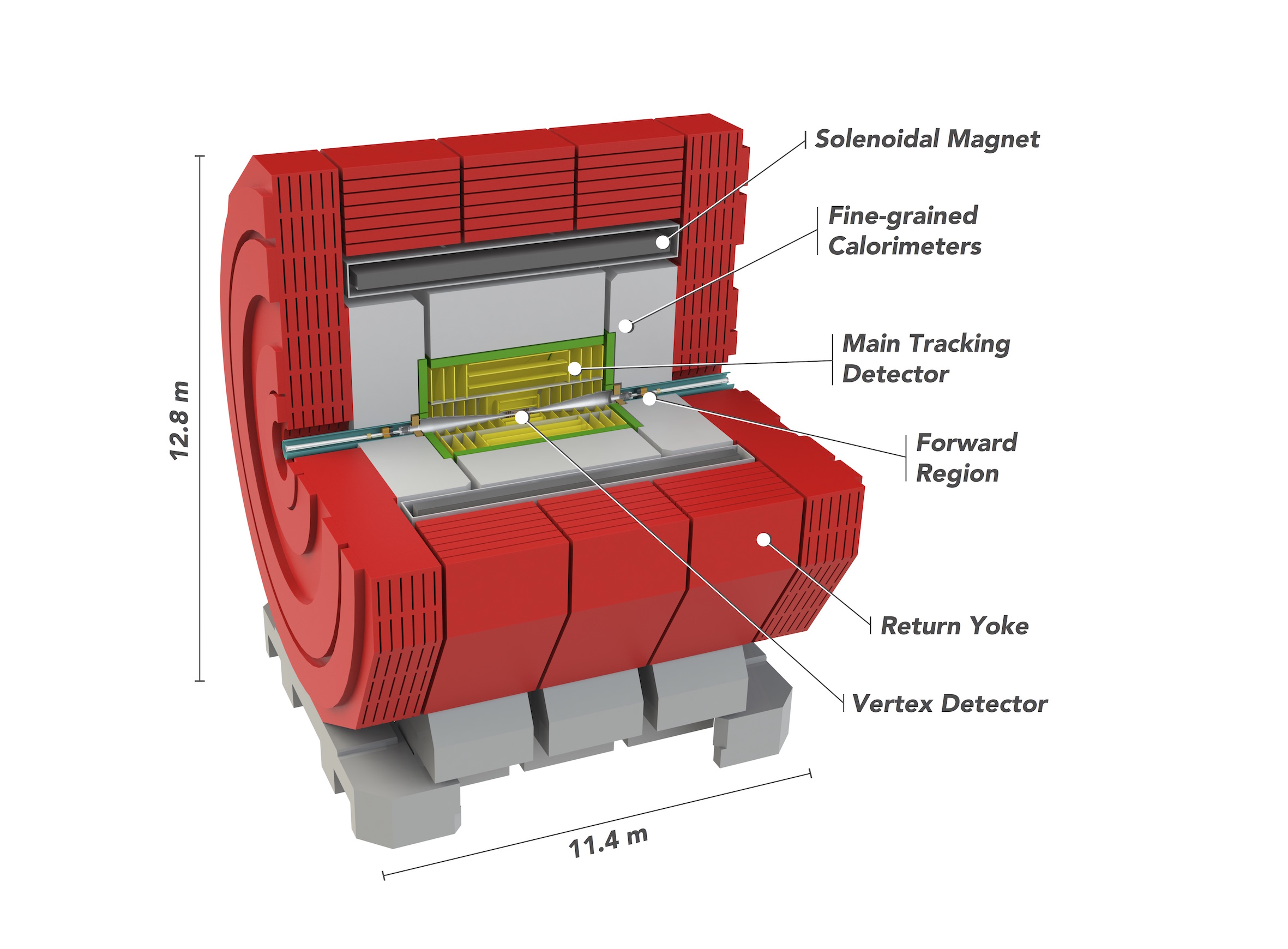}}
\end{subfigure}
\begin{subfigure}{.1075\textwidth}
\end{subfigure}
\begin{subfigure}{.37\textwidth}
  \centering
   \subcaptionbox{\label{fig:vertex_3D_view}}{  \includegraphics[width=\linewidth]{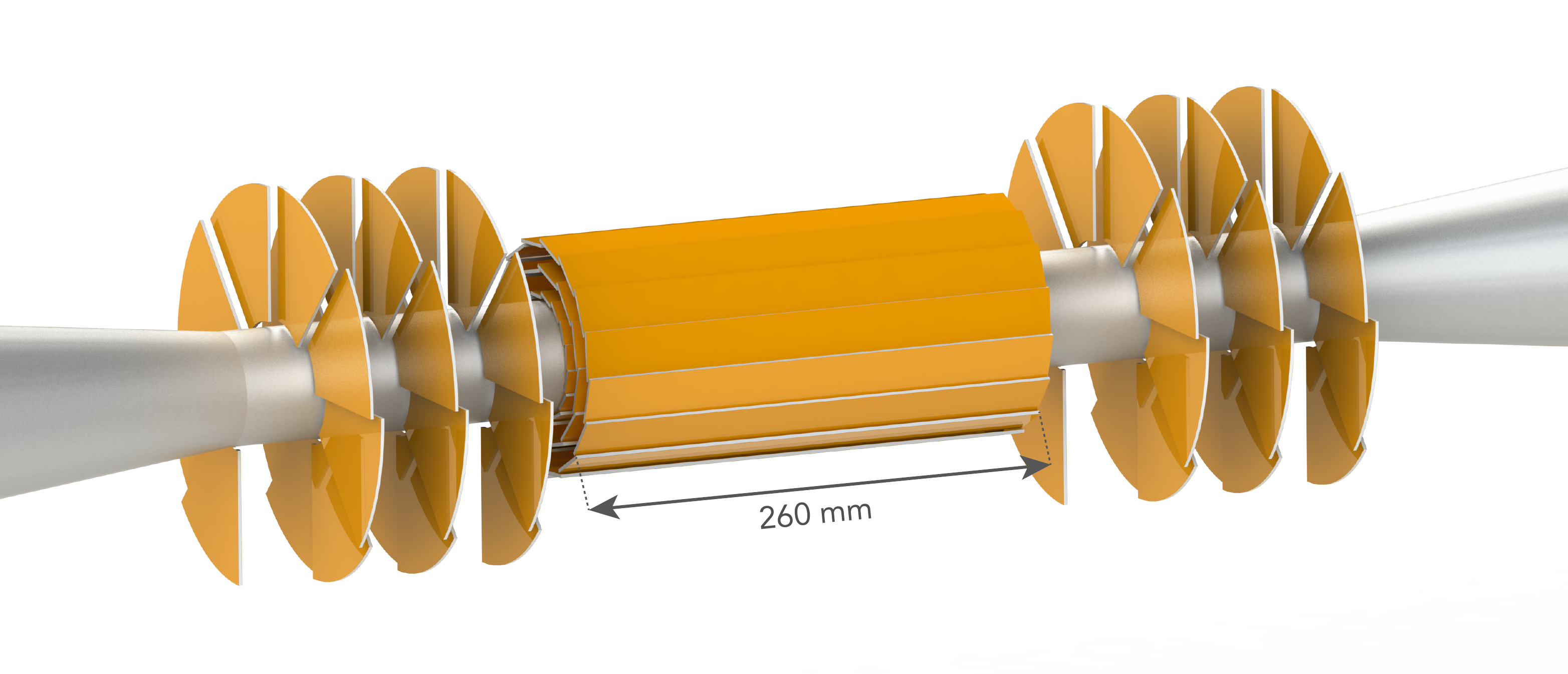}}
   \subcaptionbox{\label{fig:forward}}{  \includegraphics[width=\linewidth]{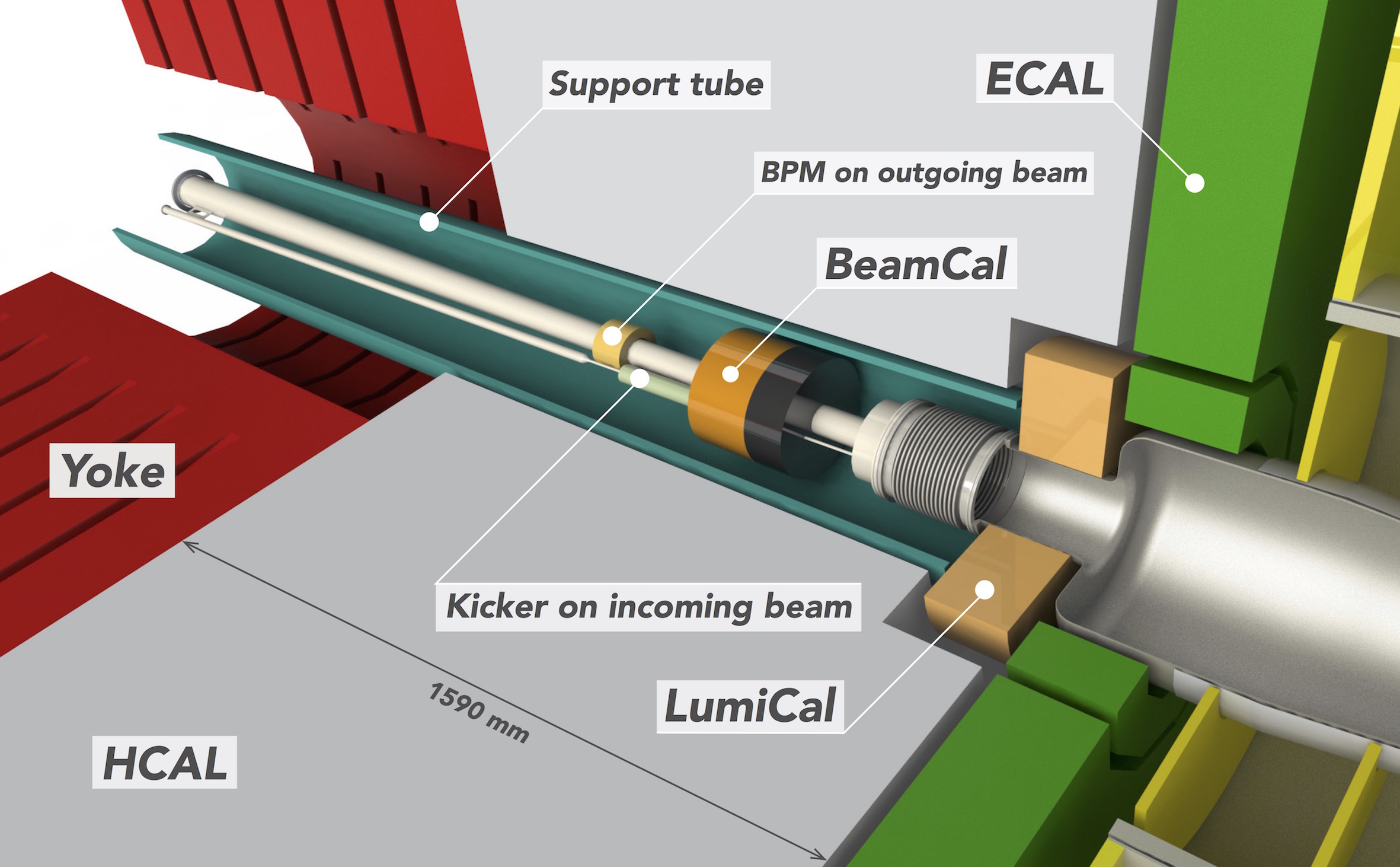}}
\end{subfigure}
\caption{(a) View of CLICdet. 
(b) View of the vertex detector layout, showing also the spiralling arrangement of the forward petals to facilitate air cooling. (c) Layout of the forward region of CLICdet. \imdp
}
\end{figure}

\paragraph{The CLIC detector}
A view of the CLIC detector (\mbox{CLICdet}) is shown in~\ref{fig:clicdet}.
It comprises a low-mass silicon-pixel vertex detector with three double layers in the central barrel
and three double layers of forward petals in a spiral arrangement optimised for air cooling.
It is surrounded by a \SI{4.5}{\meter} long, \SI{1.5}{\meter} diameter, low-mass silicon tracker,
and highly-granular calorimetry. The silicon tracker has six barrel and seven disk layers for a
total surface of \SI{140}{\square\meter}. The electromagnetic calorimeter is composed of a \num{40}-layer
sandwich with \SI{5x5}{\mm} silicon pad sensors interspersed with \SI{1.9}{\mm} tungsten plates,
for a total depth of $22\,\xo$. The hadronic calorimeter comprises \num{60} layers of plastic scintillator
tiles interspersed with \SI{19}{\mm} thick steel plates, for a total depth of $7.5\,\lambdaint$.
The scintillator tiles have \SI{3x3}{\cm} lateral size and are read out individually by silicon photomultipliers (SiPM).
The above detectors are surrounded by a superconducting solenoid providing a magnetic field of \SI{4}{\tesla}.
Detectors for muon identification are inserted into slots in the iron yoke.
The forward calorimeters, LumiCal and BeamCal, are optimised for the luminosity measurements and forward-electron tagging. The estimated data volume per bunch train is \SI{\sim115}{\mega\byte} at 3\,TeV and \SI{\sim75}{\mega\byte} at 380\,GeV.
A more detailed description of \mbox{CLICdet} can be found in~\cite{CLICdet_note_2017,CLICdet_performance}.
CLICdet was optimised for operation at $\roots=\SI{3}{\TeV}$. Since background rates at $\roots=\SI{380}{\GeV}$
are lower, some modifications to the inner detector layers are anticipated for the first energy stage~\cite{cdrvol2}.
An enlarged view of the vertex detector is shown in~\ref{fig:vertex_3D_view}.
The forward region of the detector is presented in~\ref{fig:forward}.

A significant change with respect to the CDR detector models~\cite{cdrvol2} is the location of the final focusing quadrupole QD0.
In order to enlarge the angular coverage of HCAL, QD0 is now located outside of the detector in the accelerator tunnel. 

\paragraph{Detector technologies}
\label{sec:CLICdetTechnologies}
A broad detector technology R\&D programme for CLIC is ongoing~\cite{ESU18RnD}. In view of the time scales
involved and limited resources, the development targets primarily those areas where CLIC requirements are the most
challenging: the silicon vertex and tracking system, the high-granularity ECAL and HCAL calorimeter systems,
and the forward electromagnetic calorimeters LumiCal and BeamCal. 

While the individual performance requirements for the vertex detector can be met using state-of-the art silicon / CMOS
detector technologies, the combination of all requirements is challenging and calls for new technological solutions.  
The ongoing comprehensive R\&D programme involves simulations, readout ASIC and sensor designs, the construction
and readout of small detector assemblies, laboratory tests and beam tests.
Both hybrid assemblies and depleted monolithic sensors are under study. While the R\&D programme focused initially
on the vertex detector with small pixels of \SI{25x25}{\micro\meter}, some of the technologies are now also
under consideration for the CLIC tracker. 

Given the results of the CLIC silicon vertex and tracker R\&D, together with expected advances in semiconductor
technologies worldwide, it can be expected that CLIC objectives will be achieved in a timely manner.
The following conclusions can be drawn from the results already obtained from the CLIC vertex and tracker R\&D~\cite{ESU18RnD}:

\begin{itemize}
\item Good signal-to-noise ratios have been achieved for the detection of signals from thin ($50\,\upmu\text{m}$) fully-depleted
planar sensors, sufficient for full detection efficiency and for satisfying the CLIC hit-time requirements.
\item For planar sensors, small sensor thicknesses, needed to reach the low-mass requirements, go together with small
charge sharing, limiting the achievable position resolution. Sensor designs with
enhanced charge sharing are underway and more advanced ASIC process technologies (e.g.\ 28\,nm instead of 65\,nm) offer
prospects for smaller pixel sizes and better position resolution.
\item Promising results were obtained from tests using monolithic CMOS technologies (SOI, HV-CMOS, HR-CMOS).
CLIC-specific fully integrated designs are underway. Monolithic technologies offer a potential
for high-precision performance over large surfaces, with a reduced material budget and at a lower cost.
\item Feasibility of power pulsing was demonstrated for module-size low-mass powering demonstrators,
including tests in a magnetic field. Power pulsing was also implemented successfully in hybrid ASICs and HV-CMOS
sensors for CLIC. Power consumption levels below $50\,\text{mW}/\text{cm}^2$ have been achieved.
\item Feasibility of air cooling was confirmed in a full-scale CLIC vertex detector mock-up with realistic heat loads.
\end{itemize}

A broad R\&D programme towards fine-grained calorimetry technologies is carried out in the framework of the
CALICE and FCAL collaborations, in close cooperation with CLICdp.
CALICE silicon-tungsten technology R\&D for ECAL and scintillator-steel technology R\&D with SiPM readout for HCAL
proceeded by first building `physics prototypes'. 
These devices include the core technology features and are large enough to assess their response to showers from
individual particles and to provide detailed shower data for validating the expected CLICdet jet performance
through PFA event reconstruction.  
For the 30-layer SiW ECAL the results show an energy resolution for electrons and photons at the level of $16.6\%\,\text{/}\,\sqrt{E}$
with a constant term of $1.1\%$. A 38-layer AHCAL prototype with $\sim 7600$ readout cells was the first device
to use SiPMs on a large scale. Its imaging capabilities allowed the exploitation of shower substructures,
such as using MIP tracklets for calibration purposes or using knowledge of local hit density for improving the
energy resolution through software compensation. Exposure to \SIrange{10}{80}{\GeV} pions yields an energy
resolution at the level of $57.6\%\,\text{/}\,\sqrt{E}$ with a constant term of $1.6\%$.
This result is further improved to $44.3\%\,\text{/}\,\sqrt{E}$ with a constant term of $1.8\%$ through software compensation. 

The next generation of highly-granular calorimeter prototypes are so-called `technological prototypes'.
Their design includes engineering constraints and scalability features for the construction,
which will also be needed for the final detectors. One example is the SiPM-on-tile hadronic calorimeter prototype,
comprising 38 detection layers for a total of nearly 22000 scintillator tiles. For this prototype,
construction and quality assurance processes have been optimised and automated.
Likewise, recent prototypes of the SiW ECAL integrate more of the engineering and scalability aspects,
which will be needed for an ECAL system with tens of millions of channels.

Technology development for the compact fine-grained forward electromagnetic calorimeters LumiCal and BeamCal
is pursued by the FCAL collaboration.
These detectors require a very small  Moli\`{e}re radius ($\sim$\SI{1}{cm}) and a very large dynamic range.
To this end 4- and 8-layer LumiCal prototypes were constructed and tested. A small effective Moli\`{e}re
radius of \SI{8.1 \pm 0.3}{\milli\meter} was measured using electron beams of \SI{5}{\GeV}.
Due to the large out-of-time beam-induced backgrounds in the forward region, the forward calorimeters are exposed to high
radiation levels, \eg an ionising dose of \SI{1}{MGy} per year for BeamCal. Therefore, beyond overall detector
design, BeamCal technology development is concentrating on radiation-hardness studies of sensor materials (GaAs, sapphire,
SiC and silicon diode sensors), and is yielding promising results.

\begin{figure}[htb]
\centering
\begin{subfigure}{.5\textwidth}
  \centering
  \includegraphics[width=\linewidth]{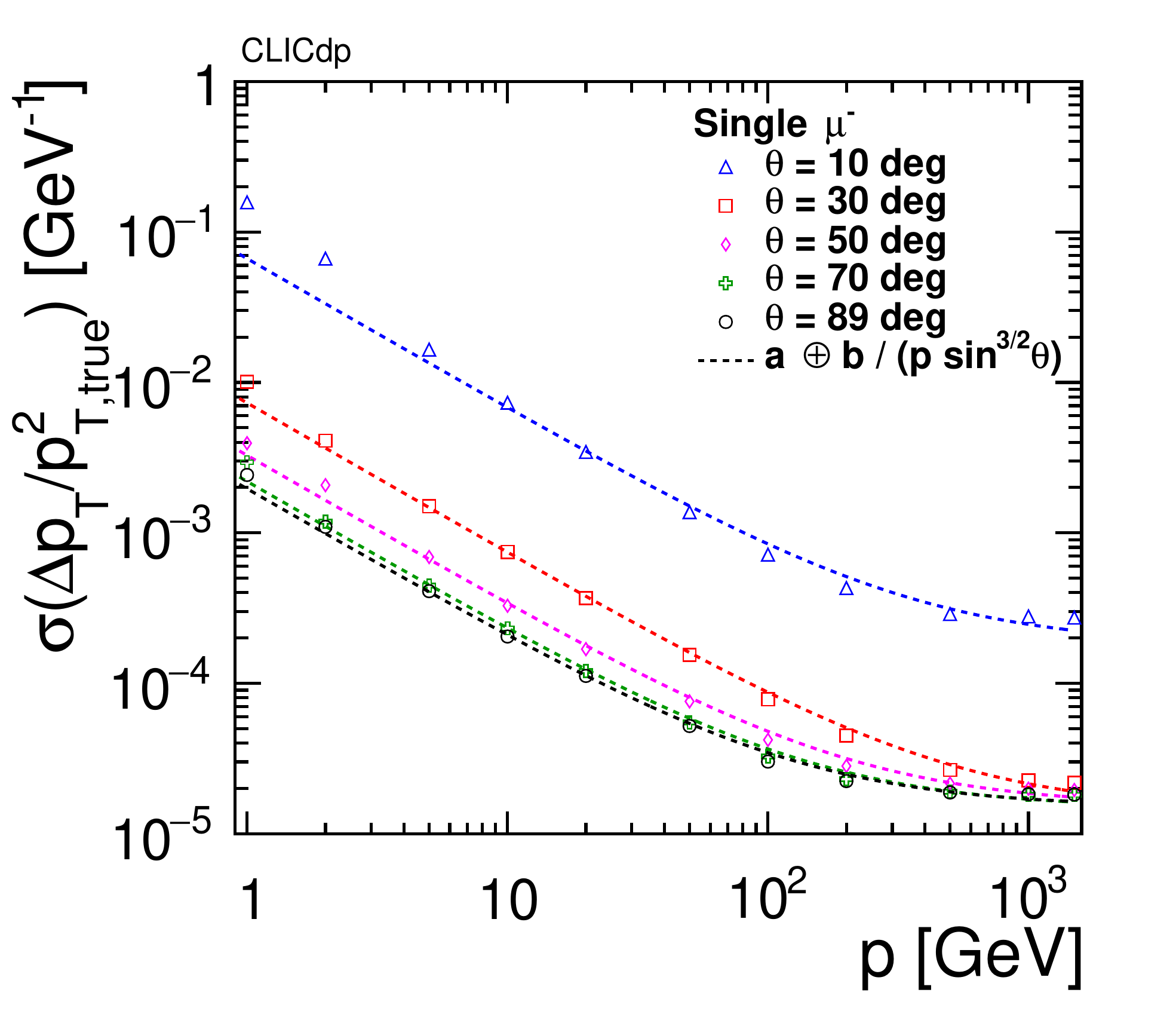}
  \caption{}
  \label{fig:mom_res_p}
\end{subfigure}%
\begin{subfigure}{.5\textwidth}
  \centering
  \includegraphics[width=\linewidth]{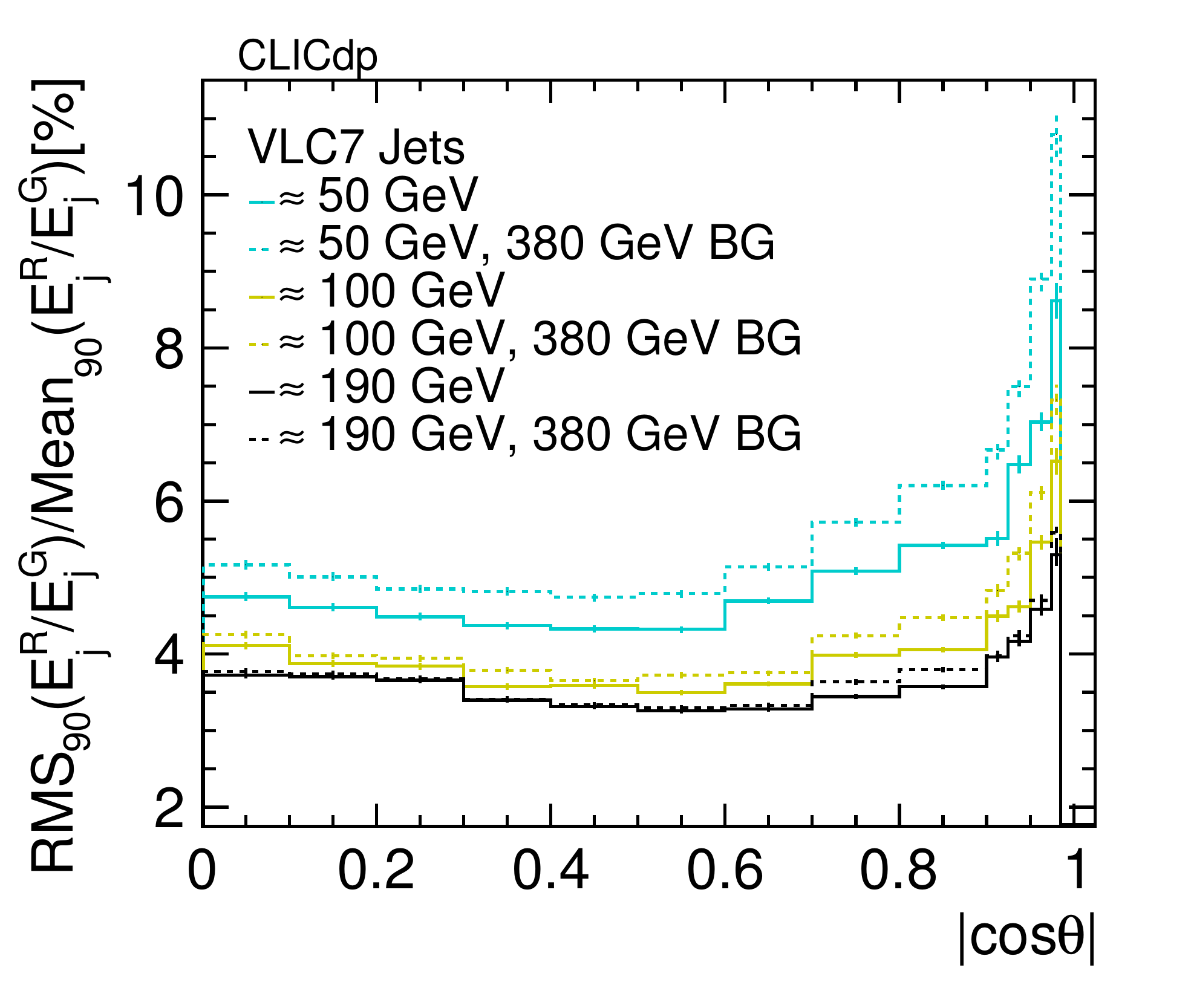}
   \caption{}
  \label{fig:jet_resp_jets_wO_380}
\end{subfigure}
\caption{(a) Transverse momentum resolution as a function of momentum for muons at polar angles \mbox{$\theta = \text{\SIlist{10;30;50;70;89}{\degree}}$}. The lines represent the fit of each curve with the parameterisation as inserted in the figure. (b)  Jet energy resolution for various jet energies as a function of $|\cos\theta|$ of the quark with and without \SI{380}{\GeV} \gghadrons background overlaid on the physics di-jet event. RMS90 is used as a measure of the jet energy resolution. \imdp}
\end{figure}

\paragraph{Detector performance}
\label{sec:CLICdetPerformance}
The CLICdet performance has been studied through simulation and event reconstruction for single particles,
complex events, and jets~\cite{CLICdet_performance}. To this end the detailed detector geometry has been implemented 
in the simulation framework; events are generated, and the resulting particles tracked through the detector. 
For example,~\ref{fig:mom_res_p} shows that the transverse momentum resolution requirement is met for
high-momentum tracks in the central detector region. The jet energy resolution
is presented in~\ref{fig:jet_resp_jets_wO_380}, showing results both for the case
without background and for the case with $\roots=\SI{380}{\GeV}$ equivalent beam-induced background overlaid and
subsequently suppressed through the timing cuts mentioned above. More examples can be found in~\cite{CLICdet_performance},
showing that the proposed CLICdet design globally meets the CLIC performance requirements.  

\begin{figure}[htb]
\centering
\begin{subfigure}{.37\textwidth}
  \centering
  \includegraphics[width=\textwidth]{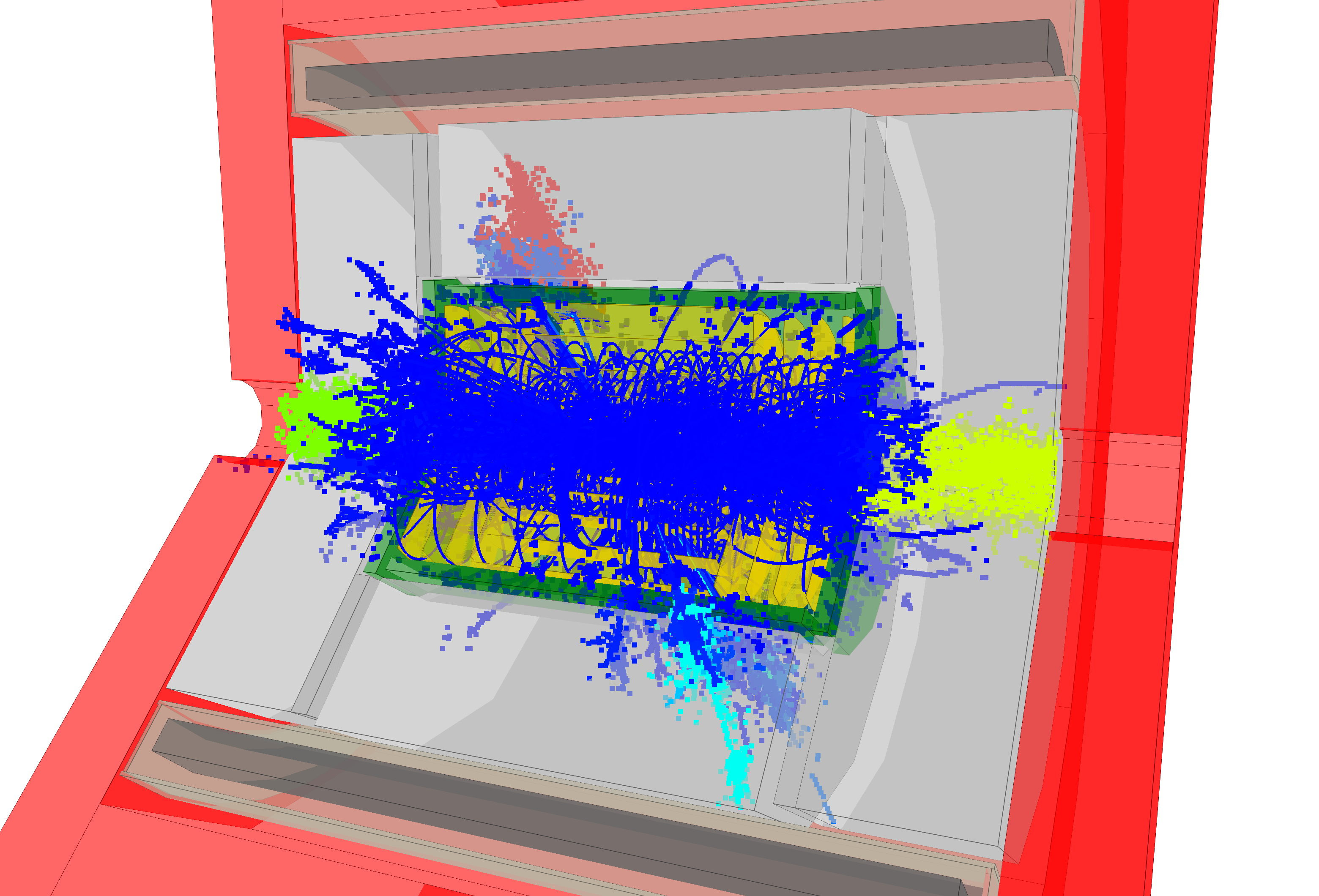}
   \caption{}
  \label{fig:tt3_before}
\end{subfigure}
\begin{subfigure}{.37\textwidth}
  \centering
  \includegraphics[width=\textwidth]{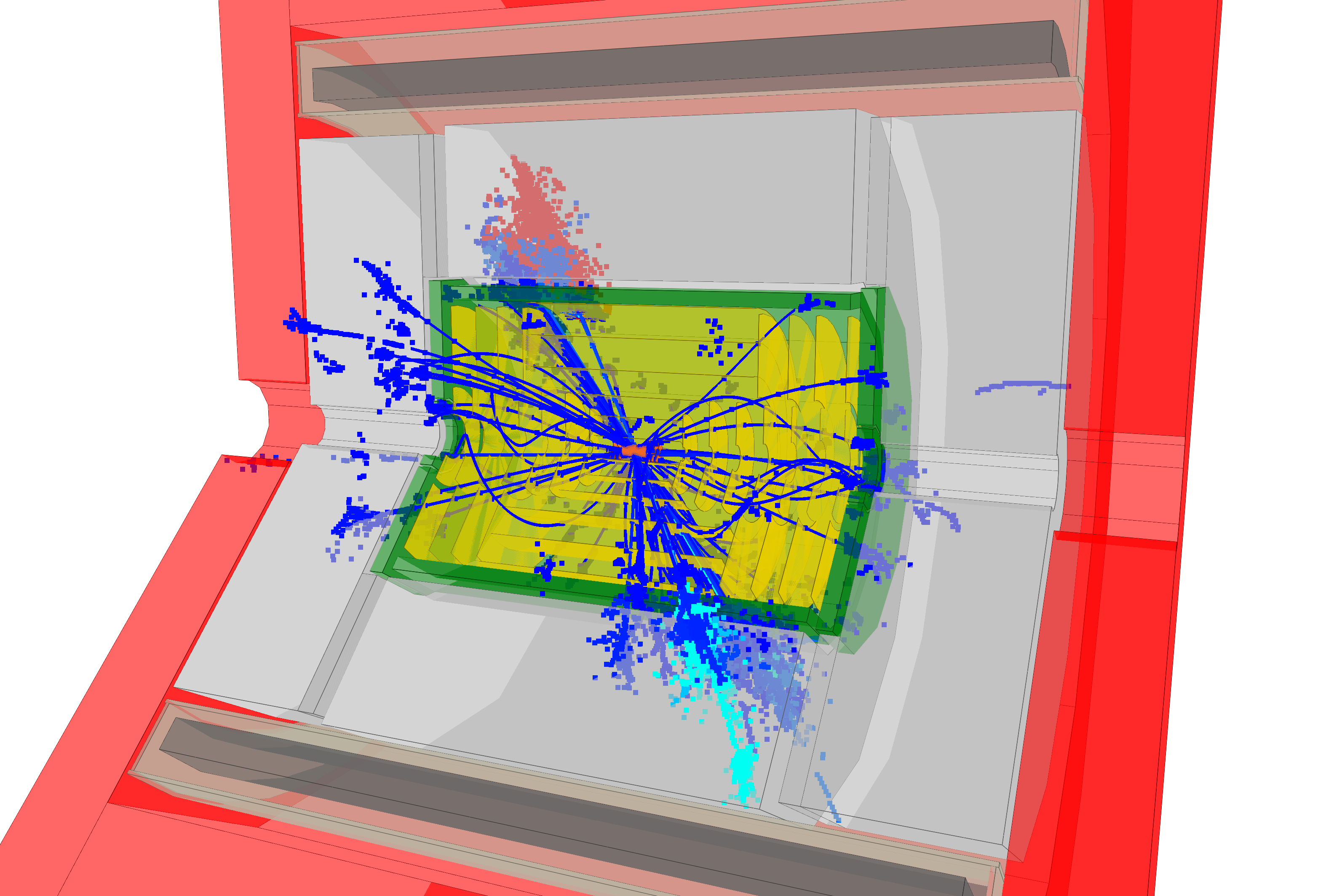}
   \caption{}
  \label{fig:tt3_After}
\end{subfigure}
\caption{$\epem\to\PQt\PAQt$ event display at a centre-of-mass energy of \SI{3}{\TeV} (a) before, and (b) after background suppression using \pT\ and timing cuts.\label{fig:tt_eventdisplay_3TeV} \imdp}
\end{figure}

To illustrate the effect of the suppression of particles from beam-induced backgrounds,~\ref{fig:tt_eventdisplay_3TeV}
depicts an event display of $\epem\to\PQt\PAQt$ at $\roots=\SI{3}{\TeV}$ before and after background
suppression using \pT\ and timing cuts. It shows that beam-induced background particles can be removed efficiently
from the event and that the resulting data show a clean boosted back-to-back $\epem\to\PQt\PAQt$ topology.

\section{Summary and Outlook}

CLIC is a mature proposal for the next generation of high-energy collider.
The choice of $\roots=\SI{380}{\GeV}$ for the initial-stage energy has been optimised for maximal physics reach,
and in turn the accelerator parameters have been systematically optimised for performance at the initial stage, 
while integrating the energy upgrade plans to \SI{1.5}{\TeV} and \SI{3}{\TeV} into the design. 
Beam experiments and hardware tests have demonstrated that the CLIC performance goals can be met.
For instance, accelerating gradients of up to \SI{145}{MV/m} have been reached with the two-beam concept at CTF3,
and breakdown rates of the accelerating structures well below the limit of $3 \times 10^7\text{m}^{-1}$ per beam pulse 
are being stably achieved at X-band test platforms.
Substantial progress has been made towards realising the nanometre-sized beams required by CLIC for high luminosities: 
the low emittances needed for the CLIC damping rings are achieved by modern synchrotron light sources;
special alignment procedures for the main linac are now available;
and sub-nanometre stabilisation of the final focus quadrupoles has been demonstrated.
Recent developments have resulted in an improved energy efficiency of the \SI{380}{\GeV} stage,
with a reduced power consumption of around \SI{170}{MW},
together with a lower estimated cost, now around \mbox{\num{6} billion \si{CHF}}.
The proposed site, at CERN, is well understood. 

The project implementation for CLIC foresees a five-year preparation phase prior to construction, which is envisaged to start by 2026. 
The preparation phase will focus on further design optimisation, technical and industrial development
of critical parts of the CLIC accelerator, and further cost and power reduction leading to a TDR around 2025.
System verification in FEL linacs and low emittance rings will be increasingly important for the CLIC accelerator performance studies.  
Furthermore, civil engineering and infrastructure preparation will become progressively more detailed, in parallel with an environmental impact study. 
The increasing use of X-band technology, either as the main RF technology or for parts of the accelerator, in \eg
compact test-facilities~\cite{Diomede:IPAC18},
light-sources~\cite{CompactLight}, medical accelerators, or for low-energy particle physics studies~\cite{Akesson:2640784}, 
provides new collaborative opportunities towards a TDR for the CLIC accelerator.

CLIC detector simulation studies have led to a new, optimised CLIC detector concept, CLICdet. 
CLICdet is optimised for particle flow with a light-weight vertex and tracking system, highly-granular calorimeter systems, 
a \SI{4}{T} solenoid and muon detectors, and 
very forward calorimeters for luminosity measurements and electron tagging.
Detector R\&D activities have validated technology demonstrators for vertex and tracking detectors and calorimetry, 
including the power pulsing of sub-detectors, and cooling concepts optimised for very thin readout layers.
For a construction start by 2026, the scope of the detector activities will need to increase significantly 
in order to cover all detectors, systems and infrastructures. Beyond the present technology demonstrators,
full prototypes of all sub-detectors, including the superconducting magnet, will be aimed for.
Detector engineering and integration aspects will be worked out in full.

The physics potential of CLIC has been assessed through many new studies and is described in a
separate submission~\cite{ESU18physics}. 
CLIC provides excellent sensitivity to Beyond Standard Model physics, through
direct searches and via a broad set of precision measurements of Standard Model processes,
particularly in the Higgs and top-quark sectors, which reach well beyond the projections for HL-LHC. 
Physics studies will continue during the preparation phase, while the development of software tools will remain
in pace with the needs of the project.
Accordingly, the CLICdp collaboration is expected to develop significantly in size and formal structure.
The current CLIC scenario spans 25--30 years of data-taking.
Longer-term, 
novel accelerator technologies~\cite{Cros2017} can potentially lead to much higher-energy beams.
The CLIC installations at CERN 
could become the home of even higher-energy \epem accelerators after completion of the CLIC programme.
Given its mature technical feasibility, outstanding physics reach, and possibility for future expansion, 
CLIC represents a compelling opportunity for the post-LHC era.

\printbibliography[title=References]
\clearpage
\appendix
\section{Addendum}

\ref{addendum:community} lists the CLIC scientific community, \ref{addendum:schedule} gives further details on the timeline,
\ref{addendum:costings} gives further details on the construction and operating costs, and \ref{addendum:computing} gives 
a short discussion of the computing requirements. 

\subsection{Community}\label{addendum:community}

The CLIC accelerator collaboration and CLIC Detector and Physics collaboration together
comprise around 400 participants from approximately 75 institutes worldwide~\cite{clic-study}.
Additional contributions are made from beyond the collaborations. 
This report is submitted on behalf of the authors of the
Compact Linear Collider (CLIC) 2018 Summary Report~\cite{ESU18Summary}:

\FloatBarrier
{\footnotesize
\begin{multicols}{2}
\begin{center}{
T.K.~Charles, 
P.J.~Giansiracusa,
T.G.~Lucas,
R.P.~Rassool,
M.~Volpi$^{1}$\\
}\textbf{University of Melbourne, Melbourne, Australia}
\end{center}
\vspace{-0.5cm}

\begin{center}{
C.~Balazs\\ 
}\textbf{Monash University, Melbourne, Australia}
\end{center}
\vspace{-0.5cm}

\begin{center}{
K.~Afanaciev, 
V.~Makarenko\\
}\textbf{Belarusian State University, Minsk, Belarus}
\end{center}
\vspace{-0.5cm}

\begin{center}{
A.~Patapenka, 
I.~Zhuk\\
}\textbf{Joint Institute for Power and Nuclear Research - Sosny, Minsk, Belarus}
\end{center}
\begin{center}
% ...
\end{center}
\vspace{-0.5cm}

\begin{flushleft}{
{$^{1}$}Also at CERN, Geneva, Switzerland\\
% ...
}
\end{flushleft}
\end{multicols}
}

\newpage
\subsection{Schedule}\label{addendum:schedule}

\begin{figure}[h!]
\centering
\includegraphics[width=0.76\textwidth]{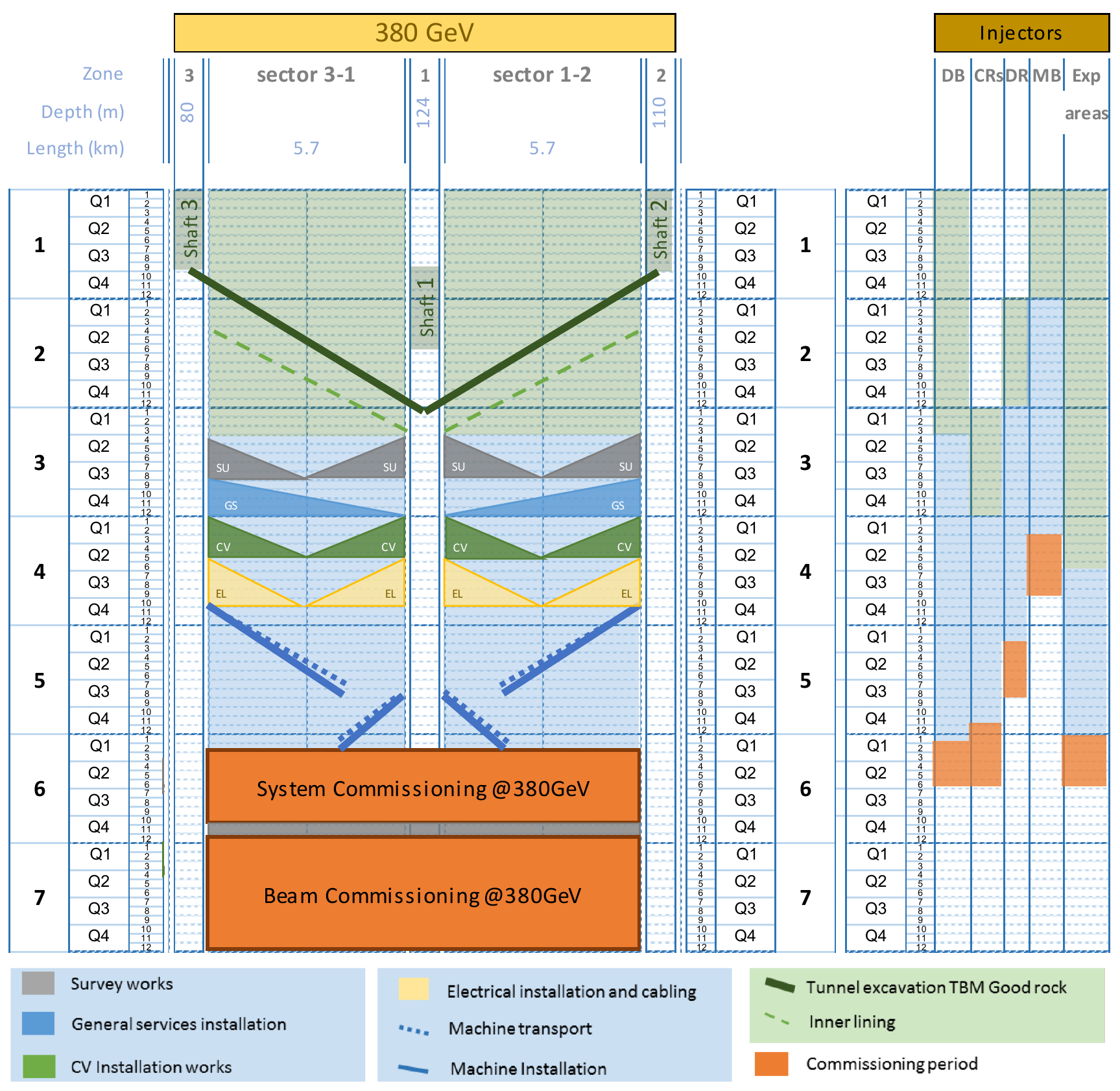}
\caption{\label{fig_IMP_6} Construction and commissioning schedule for the 380\,GeV drive-beam based CLIC facility. 
The vertical axis represents time in years. The abbreviations are introduced in~\ref{scd:clic_layout}. \imcl }
\end{figure}

The construction schedules are based on the same methodologies used for the CLIC CDR~\cite{cdrvol1}. 
Details about the various parameters used can be found in~\cite{ESU18PiP}.
The construction, installation, and commissioning schedule for the \SI{380}{\GeV} drive-beam design 
is shown in~\ref{fig_IMP_6} and lasts for 7 years. It comprises:

\begin{itemize}
\item  Slightly more than five years for the excavation and tunnel lining, the installation of the tunnel infrastructures, and the accelerator equipment transport and installation.
\item  Eight months for the system commissioning, followed by two months for final alignment.
\item  One year for the accelerator commissioning with beam.
\end{itemize}

In parallel, time and resources are allocated for the construction of the drive-beam surface building, 
the combiner rings, damping rings, main-beam building and experimental areas, and their corresponding
system installation and commissioning, as shown in~\ref{fig_IMP_6}.

In the klystron-based \SI{380}{\GeV} option, the time needed for construction, installation and commissioning is 8 years.

The \SI{380}{\GeV} collider is designed to be extended to higher energies. 
Most of the construction and installation work can be carried out in parallel with the data-taking. 
However, it is estimated that a stop of two years in accelerator operation is needed between two energy stages. 
This time is needed to make the connection between the existing machine and its extensions, 
to reconfigure the modules used at the existing stage for their use at the next stage, 
to modify the beam-delivery system, to commission the new equipment, and to commission the entire new accelerator complex with beam. 

In a schedule driven by technology and construction, the CLIC project would cover 34 years, counted from the start of construction: 
around 7 years for construction and commissioning and a total of 27 years for data-taking at the three energy stages, 
including two 2-year intervals between the stages.

\subsection{Costings and power consumption}\label{addendum:costings}

\paragraph{Construction costs}

The initial energy stage of CLIC has been fully re-costed in 2018, using the same method as
previous CLIC cost estimates and those of other projects such as the LHC experiments and the 
Reference Design Report and Technical Design Report of the International Linear Collider (ILC).
Since the CLIC CDR, many CLIC optimisation studies have been undertaken with a particular focus on cost reduction.
Further details of the cost estimate are given in~\cite{ESU18Summary} and full details in~\cite{ESU18PiP};
only a brief summary is given here.
For the cost estimate, CLIC is assumed to be a CERN-hosted project, constructed and operated
within a collaborative framework with participation and contributions from many international partners. 
A "value and explicit labour" methodology is applied, and for the value estimates 
a bottom-up approach is used that follows the work breakdown structure of the project.
Estimates are given for both the drive-beam based and klystron-based options.
The uncertainty objective for the final outcome is $\mathrm{\pm}$25\%.
Uncertainties on individual items are grouped into two categories: 
\textit{technical uncertainty}, relating to technological maturity
and likelihood of evolution in design or configuration; and 
\textit{commercial uncertainty}, relating to uncertainty in commercial procurement.

The value estimates given cover the project construction phase, from approval to start of commissioning with beam. 
They include all the domains of the CLIC complex from injectors to beam dumps, together with the corresponding civil engineering and infrastructures. Items such as specific tooling required for the production of the components, reception tests and pre-conditioning of the components, and commissioning (without beam) of the technical systems, are included.
On the other hand, items such as R\&D, prototyping and pre-industrialisation costs, acquisition of land and underground rights-of-way, computing, and general laboratory infrastructures and services (e.g.\ offices, administration, purchasing and human resources management) are excluded. Spare parts are accounted for in the operations budget. The value estimate of procured items excludes VAT, duties and similar charges, taking into account the fiscal exemptions granted to CERN as an Intergovernmental Organisation.

The CLIC value estimates are expressed in Swiss francs (CHF) of December 2018.
The breakdown of the resulting cost estimate up to the sub-domain level is presented in~\ref{Tab:Cost}.
The injectors for the main beam and drive beam production are among the most expensive parts of the project,
together with the main linac, and the civil engineering and services.

\begin{table}[ht]
\caption{Cost breakdown for the 380\,GeV stage of the CLIC accelerator, for the drive-beam baseline option and for the klystron option.}
\label{Tab:Cost}
\centering
\begin{tabular}{l l S[table-format=4.0] S[table-format=4.0]}
\toprule
\multirow{2}{*}{Domain} & \multirow{2}{*}{Sub-Domain} & \multicolumn{2}{c}{Cost [\si{MCHF}]} \\
 &  & {Drive-beam} & {Klystron} \\ \midrule
 \multirow{3}{*}{Main Beam Production} & Injectors & 175 & 175 \\
 & Damping Rings & 309 & 309 \\
 & Beam Transport & 409 & 409 \\ \hline
\multirow{3}{*}{Drive Beam Production} & Injectors & 584 &  {---} \\
 & Frequency Multiplication & 379 & {---}  \\
 & Beam Transport & 76 &  {---} \\ \hline
\multirow{2}{*}{Main Linac Modules}  & Main Linac Modules & 1329 & 895 \\
 & Post Decelerators  & 37 &  {---}  \\ \hline
Main Linac RF  & Main Linac Xband RF & {---} & 2788 \\ \hline
\multirow{3}{*}{\makecell[l]{Beam Delivery and \\ Post Collision Lines}}   & Beam Delivery Systems & 52 & 52 \\
 & Final focus, Exp. Area & 22 & 22 \\
 & Post Collision Lines / Dumps & 47 & 47 \\ \hline
Civil Engineering & Civil Engineering & 1300 & 1479 \\ \hline
\multirow{4}{*}{Infrastructure and Services}  & Electrical Distribution  & 243 & 243 \\
 & Survey and Alignment & 194 & 147 \\
 & Cooling and Ventilation  & 443 & 410 \\
 & Transport / Installation & 38 & 36 \\ \hline
\multirow{4}{*}{\makecell[l]{Machine Control, Protection \\ and Safety Systems}} & Safety System  & 72 & 114 \\
  & Machine Control Infrastructure & 146 & 131 \\
 & Machine Protection & 14 & 8 \\
 & Access Safety \& Control System & 23 & 23 \\ \midrule
\bfseries Total (rounded) & & \bfseries 5890 & \bfseries 7290 \\
\bottomrule
\end{tabular}
\end{table}

The cost estimate of the initial stage including a 1$\sigma$ overall uncertainty is:
\begin{equation*}
  \textrm{CLIC \SI{380}{\GeV} drive-beam based:} \;\; 5890^{+1470}_{-1270}\,\textrm{MCHF}; 
\end{equation*}
\begin{equation*}
  \textrm{CLIC \SI{380}{\GeV} klystron based:} \;\; 7290^{+1800}_{-1540}\,\textrm{MCHF}. 
\end{equation*}
The estimated cost of the initial stage is approximately \num{5.9}~billion~\si{CHF}.  The energy upgrade
to \SI{1.5}{\TeV} has an estimated cost of approximately \num{5.1}~billion~\si{CHF}, including the upgrade of the
drive-beam RF power.  The cost of the further energy upgrade to \SI{3}{\TeV}
has been estimated to be approximately \num{7.3}~billion~\si{CHF}, including the construction of a second
drive-beam complex.

A first estimate of the explicit labour needed for construction of the \SI{380}{\GeV} stage of the
CLIC accelerator complex was obtained
by scaling with respect to the LHC -- a CERN-hosted collider project of similar size to CLIC -- 
yielding 11500\,FTE-years of explicit labour.  This is in line with the ILC estimate of $1.8\,\text{FTE-year}/\text{MCHF}$.

The methodology used for estimating the cost of the CLIC detector~\cite{CLICdet_note_2017}
is similar to that used for the accelerator complex, and is based on the detector work breakdown structure. 
A breakdown of the value estimate for the CLIC detector is given in~\ref{Tab:Det_Cost}, and comes to a total of approximately \num{400}~million~\si{CHF}.
The main cost driver is the cost of the silicon sensors
for the 40-layer Electromagnetic Calorimeter (ECAL). For example, a 25\% reduction in the cost of silicon per unit of surface would reduce the overall detector cost by more than 10\%. Alternative designs for ECAL are feasible, but will reduce the detector performance (e.g. worse energy resolution for photons~\cite{CLICdet_note_2017}).

\begin{table}[ht]
\caption{Cost estimate of the CLIC detector~\cite{clicdet_cost}.}
\label{Tab:Det_Cost}
\centering
\pgfplotstableread[header=true]{
name z
Vertex 13
{Silicon Tracker} 43
{Electromagnetic Calorimeter} 180
{Hadronic Calorimeter} 39
{Muon System} 16
{Coil and Yoke} 95
{Other} 11
{ } {nan}
{Total } 397
}\data

\pgfplotstableset{create on use/error/.style={
    create col/expr={\thisrow{z}
    }
  }
}

\pgfplotsset{select coords between index/.style 2 args={
    x filter/.code={
        \ifnum\coordindex<#1\def\pgfmathresult{}\fi
        \ifnum\coordindex>#2\def\pgfmathresult{}\fi
    }
}}

\newcommand{\errplot}{%
\begin{tikzpicture}[trim axis left,trim axis right]
\begin{axis}[y=-\baselineskip,
xbar,
xmax=50,
width             = 7.5cm,
axis y line=none,
ytick             = \empty,
xtick={0,10,...,50},
xticklabels = {0,10\%,20\%,30\%,40\%,50\%},
axis x line*      = bottom,
]
\addplot+[xbar,fill=orange,draw=black,select coords between index={0}{7}] table [x expr ={\thisrowno{1}/397*100},y expr=\coordindex]{\data};
\end{axis}
\end{tikzpicture}
}

\pgfplotstablegetrowsof{\data}
\let\numberofrows=\pgfplotsretval

\pgfplotstabletypeset[columns={name,error,z},
  col sep = comma,
  every head row/.style = {before row=\toprule, after row=\midrule},
  every last row/.style = {after row=[0ex]\bottomrule, before row=[1ex]\midrule},
  columns/name/.style = {string type, column name=System},
      every row 7 column 2/.style={
        postproc cell content/.style={
          @cell content=\textcolor{white}{##1}
        }
      },
  columns/error/.style = {
    column name = {Cost fraction},
    assign cell content/.code = {
    \ifnum\pgfplotstablerow=0
    \pgfkeyssetvalue{/pgfplots/table/@cell content}
    {\multirow{\numberofrows}{6cm}{\errplot}}%
    \else
    \pgfkeyssetvalue{/pgfplots/table/@cell content}{}%
    \fi
    }
  },
  columns/z/.style    = {column name = {Cost [\si{MCHF}]},   column type={S[table-format=2.1]}, string type},
]{\data}

\end{table}

\paragraph{Operating costs}

A preliminary estimate of the CLIC accelerator operation cost, including replacement/operation of
accelerator hardware parts, RF systems, and cooling, ventilation, electronics, and electrical infrastructures, 
amounts to 116 MCHF per year.

An important ingredient of the operation cost is the CLIC power consumption and the corresponding energy cost. 
This is difficult to evaluate in CHF units, as energy prices are likely to evolve. The expected energy consumption
of the \SI{380}{\GeV} CLIC accelerator, operating at nominal luminosity, corresponds to 2/3 of CERN's current total energy consumption. 

The level of personnel needed for CLIC operational support is expected to be at the level of ILC estimates of 640\,FTE.

Given the considerations listed above, one can conclude that operating CLIC is well within the resources deployed for operation at CERN today. 
Operating CLIC concurrently with other programmes at CERN is also technically possible. This includes LHC, as both
accelerator complexes are independent. Building CLIC is not destructive with respect to the existing CERN accelerator complex.
Electrical grid connections are also independent. The most significant limitation will therefore be the resources,
in particular personnel and overall energy consumption.

\begin{figure}[t]
\centering
\begin{adjustbox}{width=\linewidth}
\begin{tikzpicture}[font=\sffamily,lines/.style={draw=none},scale=1,align=left]
\sansmath
\pie [
text = legend,
radius = 4.5,
sum=auto,
every only number node/.style={text=white},
style={lines},
pos={0,0},
    color={
    blue!60, 
    cyan!60, 
    yellow!60, 
    orange!60, 
    red!60, 
    blue!60!cyan!60, 
    cyan!60!yellow!60, 
    red!60!cyan!60, 
    red!60!blue!60, 
    orange!60!cyan!60 
    },
] {
1/  Main beam injectors,
1 /  Main beam damping rings,
1 /  Main beam booster and transport,
1 /  Drive beam injectors,
1 /  Drive beam frequency multiplication and transport,
1 /  Two-beam acceleration,
1 /  Main linacs (klystron),
1 /  Interaction region,
1 /  Infrastructure and services,
1 /  Controls and operations
}

\pie [rotate = 90,
radius = 4.5,
sum=auto,
every only number node/.style={text=white},
style={lines},
    color={
    blue!60, 
    cyan!60, 
    yellow!60, 
    orange!60, 
    red!60, 
    blue!60!cyan!60, 
    red!60!cyan!60, 
    red!60!blue!60, 
    orange!60!cyan!60 
    }
] {
6/  ,
53/ ,
9/ ,
45/ ,
14/ ,
3/ ,
4/ ,
33/ ,
1/ 
}
\pie [rotate = 90,
radius = 4.5,
sum=auto,
every only number node/.style={text=white},
style={lines},
pos={18,0},
    color={
    blue!60, 
    cyan!60, 
    yellow!60, 
    cyan!60!yellow!60, 
    red!60!cyan!60, 
    red!60!blue!60, 
    orange!60!cyan!60 
    }
] {
6 /,
41 /,
8 / ,
75 /,
4 / ,
29 / ,
1 / 
}
\node at (18,5) {\Large Klystron based option: 164\,MW};
\node at (0,5) {\Large Drive beam option: 168\,MW};
\end{tikzpicture}
\end{adjustbox}
\caption{\label{fig_IMP_11} Breakdown of power consumption between different domains of the CLIC accelerator in \si{\MW} at a centre-of-mass energy of \SI{380}{\GeV}, for the drive beam option on the left and for the klystron option on the right. The contributions add up to a total of \SI{168}{\MW} and \SI{164}{\MW} in the two cases. \imcl}
\end{figure}
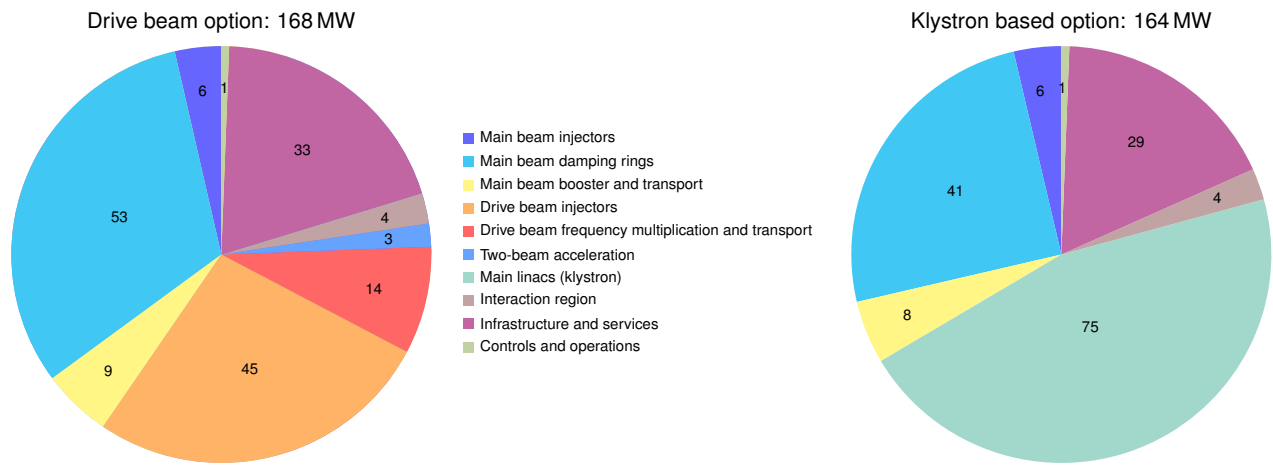

\newpage
\paragraph{Power and energy consumption}
The nominal power consumption at the \SI{380}{\GeV} stage has been estimated based on
\begin{wrapfigure}{r}{0.50\textwidth}
\begin{center}
\vspace{-10pt} \includegraphics[width=\linewidth]{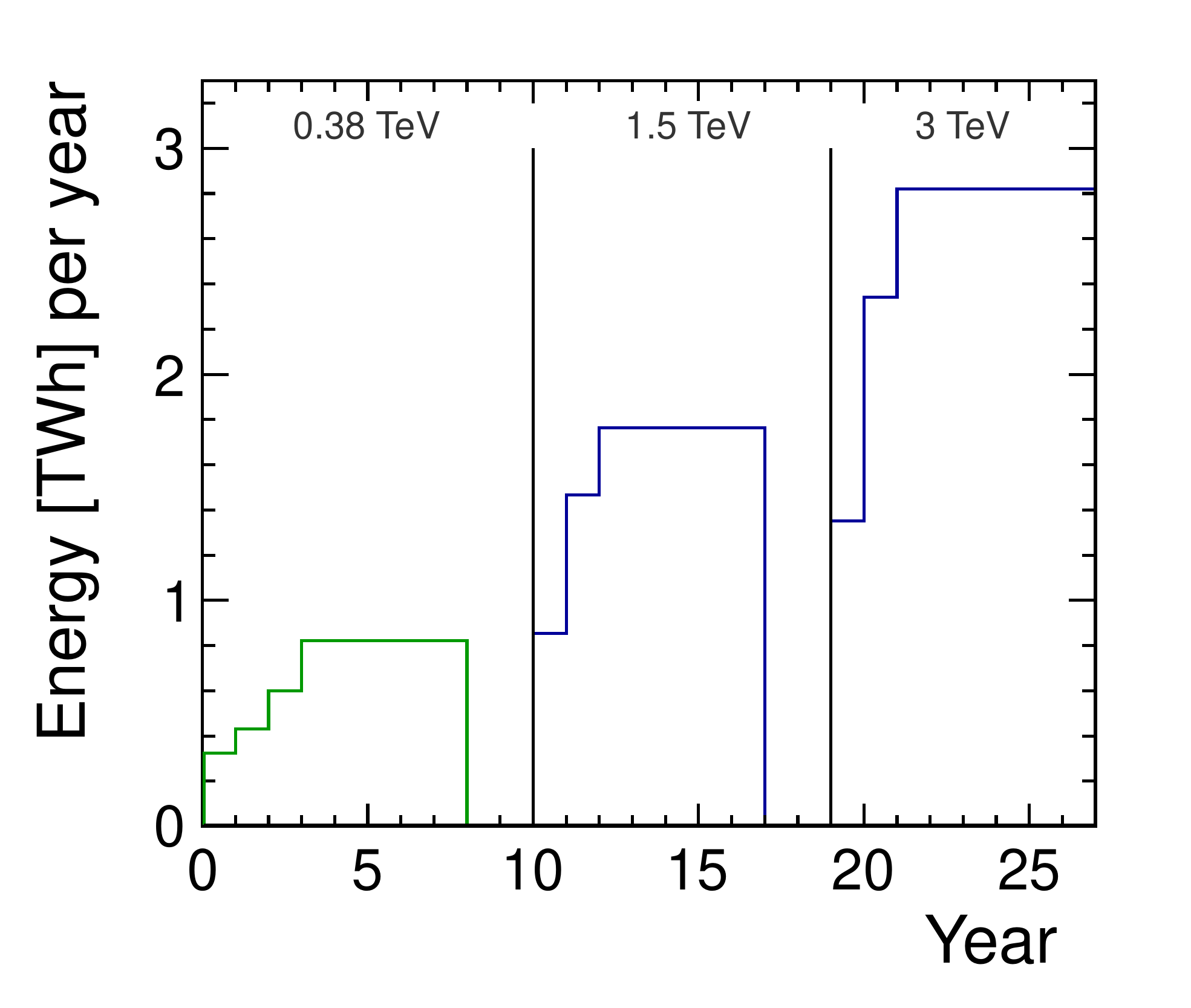}
\caption{\label{fig_IMP_13} Estimated yearly energy consumption of CLIC. The initial stage estimate is revised in detail (green), while numbers for the higher energy stages are from~\cite{cdrvol3} (blue). \imcl}
\end{center}
\vspace{-10pt}
\end{wrapfigure}
the detailed CLIC work breakdown structure.
This yields for the drive-beam option a total of \SI{168}{MW} for all accelerator
systems and services, taking into account network losses for transformation and distribution on site.
The breakdown per domain in the CLIC complex (including experimental area and detector) 
and per technical system is shown in~\ref{fig_IMP_11}.

These numbers are significantly reduced compared to earlier estimates due to optimisation of the injectors for \SI{380}{\GeV},
introducing optimised accelerating structures for this energy stage, significantly improving the RF efficiency, and consistently using
the expected operational values instead of the full equipment capacity in the estimates. 

The electrical energy consumption estimate is derived from the power consumption described above,
and the operational scenario including ramp-up.
For the \SI{1.5}{\TeV} and \SI{3.0}{\TeV} stages the recent power improvements have not been studied in detail
and the power estimates from the CDR are used~\cite{cdrvol3}. 
The time evolution of the electrical energy consumption is illustrated in~\ref{fig_IMP_13}.
For comparison, CERN's current energy consumption is approximately \SI{1.2}{TWh} per year, of which the accelerator complex uses approximately 90\%.

\subsection{Computing}\label{addendum:computing}
In view of the time scale for the project, no detailed planning of the computing needs for the detector has been established so far.
An upper limit of the data volume per train and the data rate written to tape was estimated for the full CLIC detector, including zero suppression and address encoding.
The data volume per bunch train ranges from \SI{75}{\mega\byte} at \SI{380}{\GeV} to \SI{115}{\mega\byte} at \SI{3}{\TeV}. With a bunch-train repetition rate of \SI{50}{\hertz}, this results
in data rates ranging from \SI{\sim4}{\giga\byte/\second} at \SI{380}{\GeV} to \SI{\sim6}{\giga\byte/\second} at \SI{3}{\TeV}~\cite{ESU18RnD}.
These numbers are mainly driven by the beam-induced backgrounds.

For comparison, the CMS permanent storage rate is currently at the level of \SI{\sim5}{\giga\byte/\second}, after the trigger and further online data treatment. Therefore, one can conclude that the computing needs for the CLIC experiment will not exceed those of a current multi-purpose LHC detector.


\end{document}